\documentclass[vecphys]{svmult}

\usepackage{overpic}
\usepackage{tikz}
\usepackage[detect-all]{siunitx}

\usepackage{makeidx}         
\usepackage{graphicx}        
\usepackage{multicol}        
\usepackage{cite}            
\usepackage[bottom]{footmisc}
\usepackage{epstopdf}

\usepackage{array}
\newcolumntype{L}[1]{>{\raggedright\let\newline\\\arraybackslash\hspace{0pt}}m{#1}}
\newcolumntype{C}[1]{>{\centering\let\newline\\\arraybackslash\hspace{0pt}}m{#1}}
\newcolumntype{R}[1]{>{\raggedleft\let\newline\\\arraybackslash\hspace{0pt}}m{#1}}

\begin{document}

\title*{Magnetic Imaging and Microscopy }
\titlerunning{Magnetic Imaging and Microscopy}

\author{Robert M. Reeve,  Hans-Joachim Elmers, Felix B\"uttner \and Mathias Kl\"aui}
\institute{Institute of Physics, Johannes Gutenberg University Mainz, Staudinger Weg 7, 55128 Mainz, Germany
\texttt{reeve@uni-mainz.de, elmers@uni-mainz.de, klaeui@uni-mainz.de}
\and Department of Materials Science and Engineering, Massachusetts Institute of Technology, Cambridge, Massachusetts 02139, USA \texttt{fbuettne@mit.edu}
}

\maketitle
\begin{abstract}
The magnetic domain configuration of a system reveals a wealth of information about the fundamental magnetic properties of that system and can be a critical factor in the operation of magnetic devices. Not only are the details of the domain structure strongly governed by materials parameters, but in thin-films and mesoscopic elements the geometry has an often pivotal effect, providing a convenient handle to tailor desired domain states.  Furthermore a full understanding of a system requires, in addition, investigation of the dynamic evolution of the spin-state, which is of particular importance for applications relying on e.g. the switching of magnetic elements. Here we review some of the main modern techniques for magnetic imaging, highlighting their respective advantages and limitations. The methods for imaging domain configurations and spin structures cover various spatial and temporal resolution scales and encompass those based on electron and x-ray microscopy as well as scanning probe techniques. Furthermore, away from the discipline of condensed-matter physics, magnetic effects are instrumental in a number of techniques for medical imaging, some key examples of which we also present.\end{abstract}
\noindent

\pagebreak

\pagebreak
\section{Introduction}
\label{sec:intro}
The magnetic~\index{domains}domain configuration of a system reveals a wealth of information about the fundamental magnetic properties of that system and can be a critical factor in the operation of magnetic devices~\cite{domains}. Not only are the details of the domain structure strongly governed by materials parameters, but in thin-films and mesoscopic elements the additional contributions of shape and configurational anisotropy mean that the spin configurations are qualitatively different from bulk systems and hence geometrical control provides a convenient handle to tailor desired domain states~\cite{shape}. Furthermore, more recently it is not only the magnetic domain patterns which are of interest, but also the spin configurations of the magnetic domain walls themselves since these have been proposed as functional elements in next generation memory, logic and sensing devices~\cite{racetrack, logic, sensing} and such quasiparticle-like spin textures interact differently with magnetic fields and currents depending on the detailed spin structure~\cite{field, current}. Hence high spatial resolution imaging techniques are becoming increasingly important. A full understanding of a system requires in addition investigation of the dynamic evolution of the spin-state, which is of particular importance for applications relying on, for example, the switching of magnetic elements. Since the dynamics of such systems are governed by precession frequencies and are then typically in the GHz regime, high temporal resolution time domain imaging is also highly desirable. Away from the discipline of condensed-matter physics, magnetic effects are instrumental in a number of techniques for medical imaging where the requirements and desired attributes of the methods are very different.

One of the first direct observations of magnetic domain structure was achieved in 1932 via the Bitter technique~\cite{bitter}\index{Bitter technique}. In this method the sample under investigation is covered with a fluid containing a suspension of ferromagnetic particles. Depending on the size and properties of the suspension and the sample, the particles are found to align in the magnetic stray field from the sample and the resulting pattern can be imaged with conventional optical or electron microscopy. Whilst this technique remains in use, in the intervening years a whole range of other imaging techniques have been developed. Given the wide range of length and time scales that can be relevant and due to the large range of systems that it is possible to image, there is no universal best technique and it is necessary to carefully select the most appropriate approach based on the particular requirements of a given application or experiment. Furthermore, since different techniques are sensitive to different magnetic properties such as the stray field or magnetization, it can be best to combine multiple options for a more comprehensive understanding. 

Here we review some of the main techniques currently employed for magnetic imaging. The aim is not to provide an exhaustive list of techniques but rather to give an overview of some of the most widely employed options, highlighting some of the particular considerations that must be taken into account when selecting an appropriate method, with the particular advantages and limitations of the techniques highlighted. References are provided to more in-depth discussions of particular techniques. Reviews and comparisons of multiple techniques are provided in~\cite{Hopster2004, domains, spinelect, newmag}. We divide the methods into different categories based on either the nature of the probing radiation, such as electron beam or x-ray illumination, or in the case of the scanning probe methods based on the principle of operation. Techniques based on optical illumination of the sample are  reviewed in Ref.~\cite{MOKE}. Finally at the end we briefly introduce some of the key modern magnetic imaging techniques for medical applications.
\section{Electron Microscopy}
\label{sec:electron}
~\index{electron microscopy}
The first class of microscopy techniques that will be discussed use electron beams in order to probe the sample. The incident electron beam can interact with the sample based on different mechanisms. In the first instance the beam may be deflected depending on the magnetic configuration, yet additionally the excitation of the sample generates secondary electrons which also carry information about the magnetic state. Due to the mature technology in  generating highly focussed electron beams, scanning approaches can offer very good spatial resolution imaging. Electron microscopy techniques tend to be limited to conducting specimens, since otherwise the sample becomes charged by the electron beam, leading to unwanted deflections and distortions. This can sometimes be overcome if insulating systems want to be investigated by coating the surface with a thin conductive layer to help mitigate charge build-up. Furthermore the application of magnetic fields during imaging is often severely limited due to their deflecting and depolarizing effects on the electrons, although there are sometimes strategies to partially overcome this~\cite{SEMPAfield}. Here we focus on methods which use an unpolarized electron beam as the probe, however we note that certain specialized approaches such as spin-polarized low energy electron microscopy employ polarized electron beams to excite the sample, details of which can be found in~\cite{emicro, Hopster2004}. 
\subsection{Transmission Electron Microscopy}
\label{subsec:TEM}
~\index{transmission electron microscopy}
Following the Bitter technique, transmission electron microscopy (TEM) was one of the earliest techniques for revealing magnetic domain structure~\cite{earlyTEM}. In TEM imaging the electron beam incident on the specimen is accelerated by high voltages, resulting in highly energetic electrons with typical energies of 100-200 keV and in some cases up to 1000 keV. The electron intensity is then detected in transmission. One key advantage of TEM is the ease of carrying out complementary non-magnetic characterization of samples in order to correlate the observed magnetic configurations with the local electronic and structural properties. However, since the signal is measured in transmission this places considerable constraints on the sample thickness, which is typically limited to about 100\,nm or less. For bulk samples it is necessary to apply thinning processes before imaging is possible, which potentially modifies the domain structure of interest. Thin film samples and associated lithographically defined nanostructures are more readily imaged via TEM, but need to be deposited on suitable substrates which are transmissive for the electrons, e.g. silicon nitride (Si$_3$N$_4$) membranes, which take some care in handling. Another important consideration when applying TEM imaging to magnetic structures is that usually the sample would be subject to a very strong magnetic field from the objective lens of the microscope. To avoid perturbing the domain structure, strategies have to be employed to reduce this field which may require dedicated equipment and in all cases tends to limit the resolution of magnetic modes of TEM microscopy as compared to other forms of TEM characterization. The existence of structural contrast even in magnetic imaging modes can also limit the practical resolution. A review of the application of TEM to imaging magnetic microstructure is provided in~\cite{TEMrev}. In the following we describe some of the different operational modes of the technique. 
\subsubsection{Lorentz Microscopy}
~\index{Lorentz microscopy}
Lorentz microscopy relies on the perturbation of an electron beam due to magnetic fields. The resulting small angular deflections of the beam of around $10^{-5}-10^{-4}$\,rad, can be classically attributed to the so-called Lorentz force:
\begin{equation}
\label{eq:lor}
\vec{f_{lor}}= \vert e\vert \,(\vec{\upsilon} \times \vec{B}) \ts,
\end{equation}
 where e is the electron charge, $\vec{\upsilon}$ the electron velocity, which depends on the acceleration energy and $\vec{B}$ is the magnetic flux density. In quantum mechanical terms the sample can be considered to modulate the phase of the incident electron wave depending on the magnetic state, leading to bright-dark contrast due to interference. From equation~\ref{eq:lor} it can be seen that components of $\vec{B}$ aligned with the beam do not contribute to the deflection and therefore samples may need to be tilted in the case of perpendicular magnetic anisotropy systems. The technique is sensitive to magnetic flux along the whole path of the electron beam and hence it is not only the magnetization within the sample which contributes to the deflection but also stray magnetic fields. In some cases these two contributions can act against each other, diminishing or even cancelling out contrast.  

\paragraph{Fresnel Imaging} %
~\index{Fresnel imaging}
The first mode of Lorentz imaging is the Fresnel imaging, or defocus mode. Since the influence of the magnetic structure of the specimen only causes deflections of the beam, an in-focus image of the sample normally does not contain any magnetic contrast. In Fresnel imaging this is overcome by defocusing the objective lens. This reveals magnetic features of the sample, however, at the expense of reducing the achievable spatial resolution. 

\begin{figure}[t]
\centering

\includegraphics*[width=.7\textwidth]{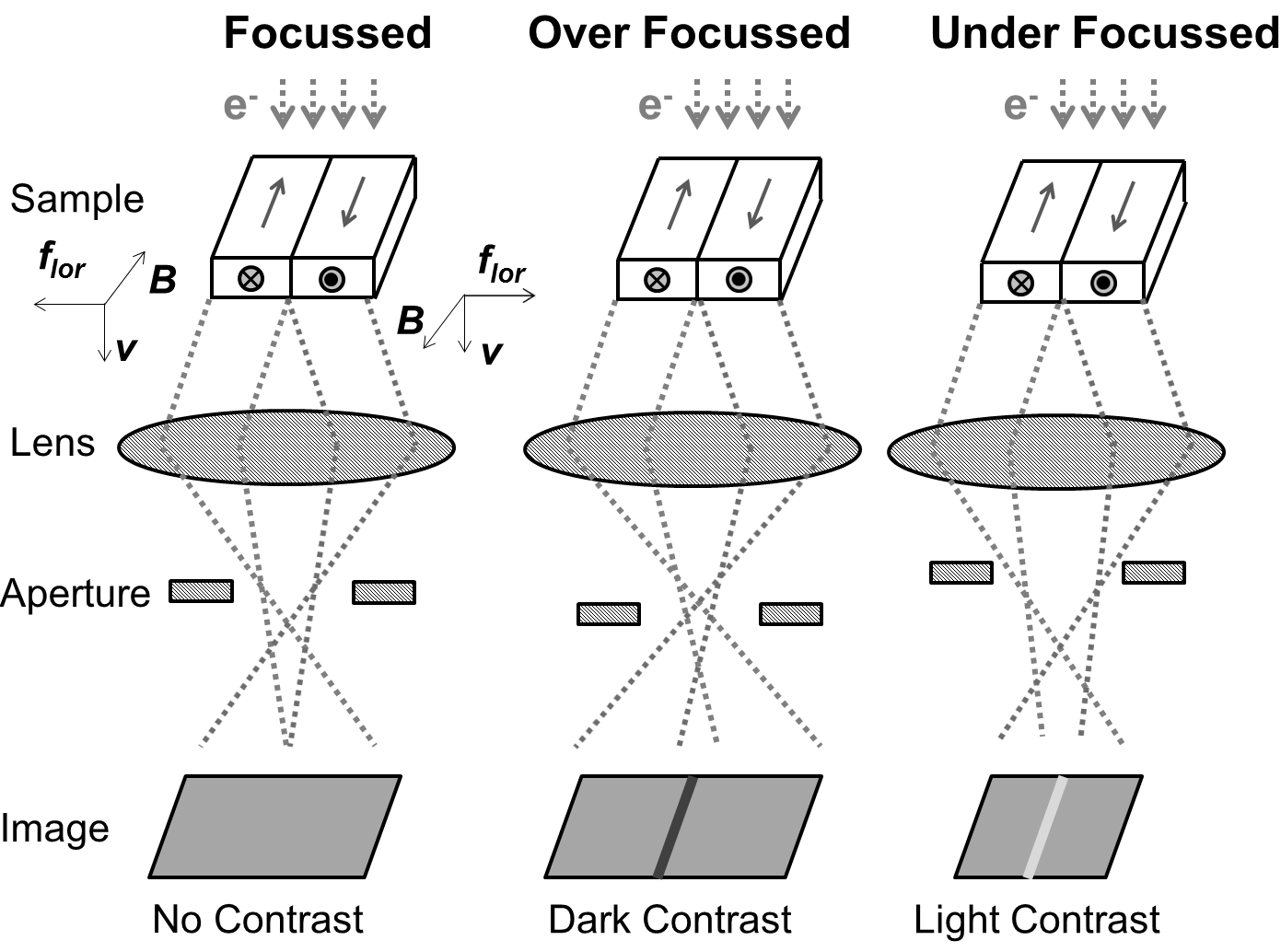}
\caption[]{Fresnel mode of Lorentz microscopy: Schematic representation of the contrast formation in the Fresnel imaging mode for a sample consisting of two opposing in-plane domains separated by a 180$^\circ$ domain wall. The Lorentz force deflects the electron beams, leading to dark/bright contrast for over- and under- focus conditions of the microscope, respectively}
\label{fig:FreTEM}     
\end{figure}

The mechanism is illustrated in Figure~\ref{fig:FreTEM}.
As can be seen in the figure, the opposite direction of deflection for neighbouring beams on either side of a 180$^\circ$ domain wall leads to either converging or diverging beams at the wall positions. Consequently such walls are revealed by corresponding bright or dark contrast, depending on whether overfocussed or underfocussed imaging is employed, whilst the domains themselves usually have uniform contrast. An exception is the case of polycrystalline films where due to small fluctuations in the directions of magnetocrystalline anisotropy characteristic ripple contrast can occur which is oriented perpendicularly to the magnetization direction of the given domain. More in-depth reconstructions of the domain states of a sample are, however, extremely challenging in this imaging mode. Due to the required high defocussing there is a very strong non-linearity between the contrast and the magnetic state of the film. More recently the possibility to reconstruct  the phase of the emerging electron wave has been demonstrated by acquiring multiple Fresnel images for different values of defocus and then applying the transport of intensity equations~\cite{phase}.
  
\paragraph{Foucalt Imaging} %
~\index{Foucalt imaging}
In Foucalt imaging a different strategy is employed to reveal magnetic contrast in TEM images~\cite{Foucalt}. Since the variously magnetized domains lead to different Lorentz deflection angles of the beams, the reciprocal space image of the sample is split into different components for these different Lorentz angles. Particular components can therefore be selected by using an aperture to block part of this reciprocal space pattern which is formed in the back focal plane of the microscope. 
\begin{figure}[t]
\centering

\includegraphics*[width=.7\textwidth]{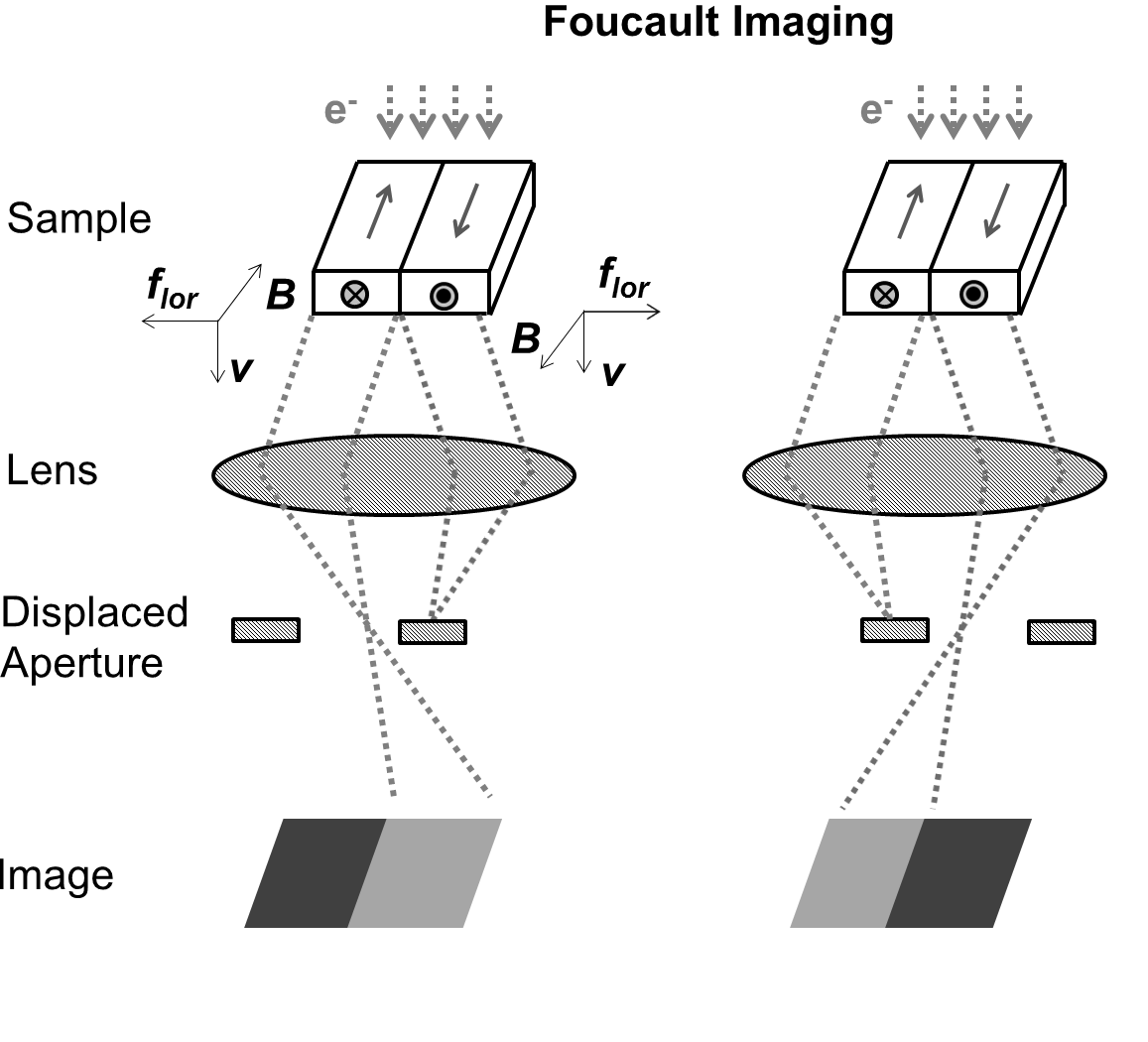}
\caption[]{Foucault mode of Lorentz microscopy: Schematic representation of the contrast formation in the Foucault imaging mode for a sample consisting of two opposing in-plane domains separated by a 180$^\circ$ domain wall. Part of the diffraction plane is blocked, leading to bright-dark contrast corresponding to the domain regions}
\label{fig:FouTEM}       
\end{figure}
This is illustrated schematically in figure~\ref{fig:FouTEM} where two spatially separated diffraction spots are evident due to the two magnetization directions present in the specimen. By blocking one or other of these beams contrast is generated in the image. Unlike in Fresnel imaging, in the Foucalt mode the contrast is correlated to the magnetic induction from the domains themselves and not the change in magnetic induction between domains. However, the stringent requirements on the quality and positioning of the aperture mean that this mode is difficult to implement.
\paragraph{Differential Phase Contrast Microscopy} %
~\index{differential phase contrast microscopy}
For more quantitative imaging, the differential phase contrast technique is an attractive option based on scanning TEM~\cite{diffphase, diffphaseorig}. The incident focussed beam is rastered across the sample and the transmitted beam detected by a special four-quadrant circular detector. For magnetic samples the Lorentz deflection leads to opposite quadrants being illuminated to a greater or lesser extent and hence difference signals for the two opposing quadrant pairs provide quantitative information of the components of the two orthogonal in-plane magnetic induction components.

This method suffers from longer recording times than the previous modes due to the necessity to scan the sample and furthermore has increased instrumental and experimental complexity. However good spatial resolution can be achieved from the focussed beam, down to around 5\,nm.

\subsubsection{Electron Holography}
~\index{electron holography}
Whereas the modes discussed so-far can conceptually be understood from a classical picture of the electron beam, electron holography explicitly relies on the quantum mechanical wave nature of the electrons. Such techniques can provide high spatial resolution imaging typically down to about 5\,nm. A wide variety of schemes exist, including even tomography, as reviewed in Refs.~\cite{holorev, twentyhol, tomography}, however, the most common mode is off-axis holography as outlined below.

\begin{figure}[t]
\centering

\includegraphics*[width=.4\textwidth]{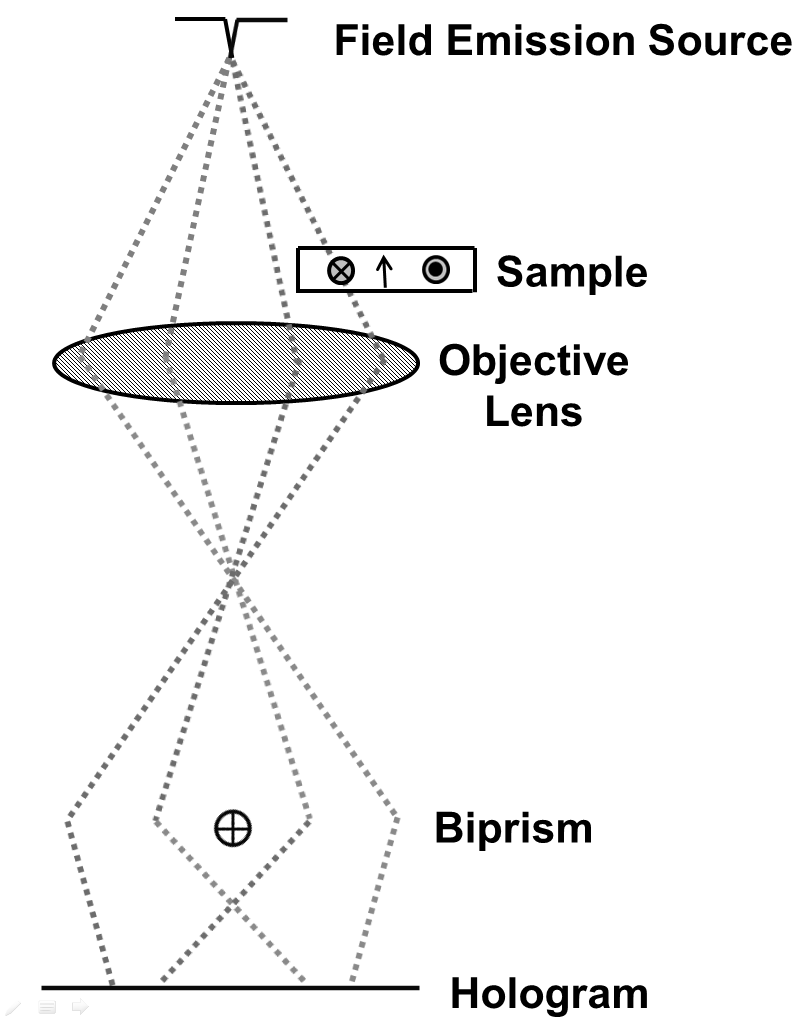}
\caption[]{Electron Holography: Schematic representation of the off-axis electron holography technique}
\label{fig:holo}       
\end{figure}

\paragraph{Off-axis Holography}
~\index{off-axis holography}In off-axis holography a highly coherent incident beam is split into a probe beam and a reference beam, the first of which passes through the sample while the latter remains unperturbed. Due to the interaction of the probe beam with the magnetic state, the electron wave acquires a phase shift depending on its path. When the probe and reference beams are then recombined they interfere to form a holographic interference image, encoding information on both the phase and amplitude of the transmitted wave. In the ideal case this is directly related to the magnetic state of the sample, however complications arise for samples with non-uniform composition or thickness since these can introduce other sources of phase shifts, for instance electrical. Quantitative information can then be extracted by processing the interference pattern to mathematically reconstruct the amplitude and phase. There are two main imaging modes. The setup for the absolute mode is depicted in Figure~\ref{fig:holo}. The sample is chosen such that it only partly fills the image plane, for example by imaging the edge of a structure or a small element. Part of the beam then passes through the specimen and part is unperturbed. In order to recombine the two beams and form the hologram an element called a biprism is used which consists of a thin metallic wire or quartz fiber coated with Au or Pt and which is biased at a voltage of typically 50-200\,V.  In the differential mode two beams are created which are both directed towards the sample, separated by a small distance and the phase shift between these two beams is then recorded. This approach is advantageous for the investigation of fine structure such as the profiles of magnetic domain walls with the resolution set by the beam diameter.

\subsubsection{Aberration Correction}
~\index{aberration correction}
In conventional TEM, there has been significant recent improvement in achievable resolution by implementing aberration correction~\cite{aberrev}. Electron lenses are inherently much poorer than optical lenses and their associated spherical aberration is often a key limiting factor in determining the resolution of electron microscopes. To counteract this, schemes have been developed to compensate for the aberration of the TEM objective lens by incorporating a correcting element with negative spherical aberration into the microscope. Two approaches are based on multipole lenses called the quadrupole-octupole corrector and sextupole corrector. In the case of magnetic imaging the objective lens is often not used due to the impact of the associated magnetic field on the sample, however aberration correction schemes can still be employed to compensate for the relevant lens in the instrument and improve the attainable resolution of these imaging modes~\cite{aber}.

\subsection{Scanning Electron Microscopy}
~\index{scanning electron microscopy}
\label{subsec:SEM}
In a scanning electron microscope (SEM) the electron beam is scanned across the sample and the generated electrons are detected in reflection. When an energetic primary electron beam interacts with a sample a spectrum of energies for electrons leaving the sample results. At high energies around the primary beam energy there is a peak corresponding to elastically scattered electrons. In the middle of the spectrum small elemental specific peaks are found corresponding to Auger electrons which are typically in the 100-2000$\,$eV range. Finally at very low energies, below around 50$\,$eV, one finds the so-called true secondary electrons. These correspond to electrons having undergone many inelastic scattering processes which are emitted in a cascade process. Since this scattering involves states in the vicinity of the Fermi level the emitted secondaries are found to be spin-polarized in itinerant ferromagnets due to the imbalance between spin up and spin down states~\cite{chrobok}. For energies above $\sim$10\,eV this spin polarization directly reflects that at the Fermi level, whereas at lowest energies an additional enhancement in polarization is observed due to a spin dependent scattering induced spin filtering effect.
The interaction of the electrons with the sample leads to different effects depending on their energy which can be employed for magnetic imaging~\cite{SEMmodes}. In the first case, as with many of the TEM techniques, the trajectories of the detected electrons are modified by the magnetic configuration of the sample.  Secondly the emitted electrons may be spin-polarized, with the polarization representing features of the spin-split band structure of the ferromagnet. If the low energy secondary electrons are detected, the contrast is termed ``type I". In this case deflection is largely due to the magnetic stray fields from the sample. The elastically backscattered electrons are primarily affected by the magnetization within the sample, leading to so-called ``type II" contrast. In this case the sample is tilted with respect to the beam and the resulting deflection of the electrons within the sample leads to an enhancement or reduction of the backscattered electron yield depending on whether the deflection is directed towards or away from the surface. While the lateral resolution, at a few $\mu$m, is quite limited compared to other magnetic imaging techniques, the deep penetration of highly energetic electrons can be taken advantage of to image through surface layers and can probe domain structures to depths of $\sim$1-20\,$\mu$m depending on the incident electron beam energy. 

\subsubsection{SEMPA}
\label{subsec:SEMPA}
\begin{figure}[t]
\centering

\includegraphics*[width=.9\textwidth]{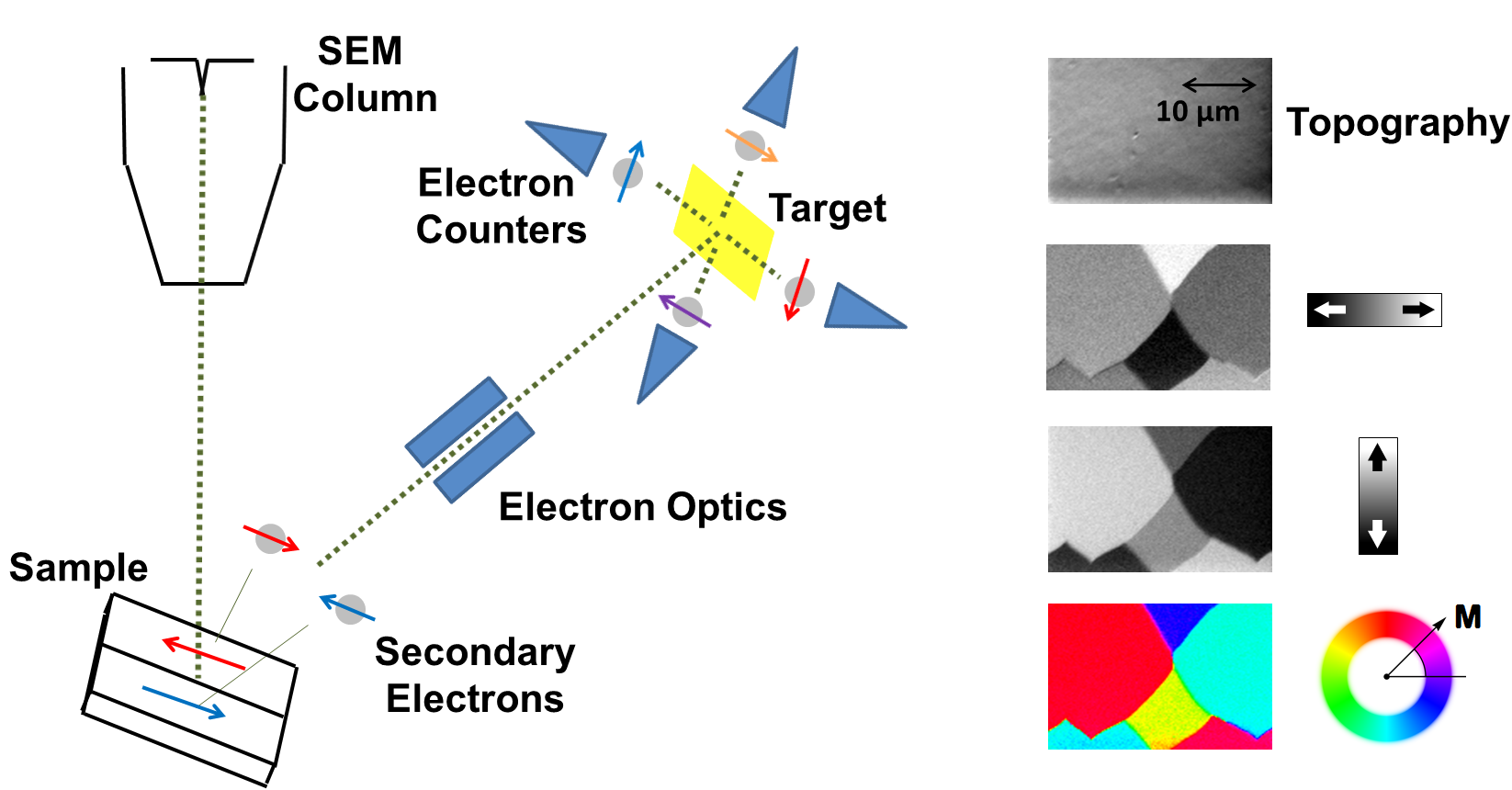}
\caption[]{SEMPA: Schematic representation of the SEMPA technique. An unpolarized SEM beam excites secondary electrons from the magnetic sample and their spin polarization is analyzed with a spin detector. On the right typical images are presented for an Fe whisker. Here the technique simultaneously acquires the topography and the two in-plane magnetization components which can be processed to provide the full in-plane colour map of the magnetization.}
\label{fig:SEMPA}       
\end{figure}

~\index{SEMPA}
~\index{spin SEM}
Scanning electron microscopy with polarization analysis (SEMPA) or spin-SEM takes advantage of the fact that the spin polarization of the emitted low-energy electrons is oriented antiparallel to the magnetization in the sample~\cite{SEMPA1, SEMPA2, SEMPA3}. Hence by exciting secondary electrons point-by-point with an unpolarized scanning electron beam and measuring the polarization of the emitted electrons, a direct quantitative representation of the domain state can be obtained as illustrated in Figure~\ref{fig:SEMPA}. 
The polarization of the emitted electrons is measured via a spin detector. Here the spin-polarized electrons are focussed onto a target where, due to the spin-orbit interaction, asymmetries in scattering exist for spin up and spin down electrons. By counting the number of electrons scattered in opposing directions using electron multipliers, the beam polarization can be calculated as follows:
\begin{equation}
\label{eq:pol}
P=\frac{1}{S_{\text{eff}}} \frac{N_A-N_B}{N_A+N_B},
\end{equation}where $S_{\text{eff}}$ is the effective Sherman or sensitivity factor of the detector which quantifies the scattering asymmetry that is obtained for a 100\% polarized beam and $N_{A/B}$ are the electron counts for scattering in opposite directions. Instruments are usually equipped with two pairs of electron counters which simultaneously provide the two orthogonal in-plane components of the magnetization which can be combined into a single map of the 2D domain structure as shown in Figure~\ref{fig:SEMPA}. Some instruments also employ a spin-rotator, in which case the out-of-plane component can also be accessed. The sum of the signals from all four detectors provides a secondary electron topographic image of the sample which is helpful in distinguishing between features of magnetic and non-magnetic origin. One advantage of SEMPA is that in general morphological details are suppressed in the magnetic asymmetry images.

A number of designs of spin detectors exist~\cite{suga_photoelectron_2015}, however the two most commonly employed in SEMPA are the Mott polarimeter and spin-polarized low energy electron diffraction (SPLEED) detector.~\index{spin detector} The SPLEED detector takes advantage of the spin dependent low-energy electron diffraction from a W(100) crystal with the asymmetries in the intensities of the (2,0) diffraction beams at 104.5 eV scattering energy employed~\cite{Fro2}. The Mott detector is based on the spin dependent Mott-scattering of highly energetic electrons from films of high atomic number elements. Originally Mott polarimeters worked at particularly high voltages and were very bulky. Nowadays, however, much more compact instruments are available operating around 29\,kV~\cite{dunning}, facilitating their employment in a small lab setting~\cite{reeve2}. Unfortunately, the inherent low efficiency of spin detectors of around $10^{-4}$ means that long acquisition times are required to obtain sufficient signal to noise ratio per pixel, with typical images requiring several minutes or longer, depending on the desired resolution, imaging area and the particular material. The entire system therefore needs to be stable on these timescales including the incident beam, sample state and mechanical vibrations. Furthermore thermal drifts are often problematic and can limit practical resolutions. In the case of the SPLEED detector the integrity of the W surface also needs to be maintained since even residual gas adsorption will degrade the performance~\cite{Fro1}, requiring that the surface is periodically flash-heated to regenerate good scattering conditions.

Since the magnetic probing depth of spin-polarized electron spectroscopies is very small~\cite{magprobe} SEMPA has an extreme surface sensitivity of around 1$\,$nm. On the one hand, this places particularly stringent requirements on the cleanliness of surfaces being measured, requiring measurement in ultra-high vacuum (UHV). For thin films, capping layers or non-magnetic oxide surface regions need to be removed in-situ by, e.g. Ar$^+$ ion sputtering while bulk crystals are often cleaved in-situ to reveal a pristine surface. For simple $3d$ metals, such in-situ sputtering often yields good results, yet for more complex compounds care needs to be taken to ensure that different sputter rates for the different elemental components do not lead to changes in the stoichiometry. An often employed strategy to mitigate the surface requirements involves the in-situ deposition of a thin dusting layer of Fe~\cite{dusting}. The expectation is that this will couple to the magnetic structure of the underlying film, imprinting the domain structure in the pristine Fe layer which can itself then be imaged. This method can also be employed to improve the contrast for materials with low signals such as non-itinerant ferromagnets and can also facilitate the imaging of insulating systems where charging effects usually prevent investigation with electron beams. However it is necessary to confirm that the thickness of the deposited Fe film is thin enough that it does not change the domain state in the material under investigation. At the same time, the surface sensitivity confers the ability to selectively probe the properties of the surface~\cite{LSMO}, which are often instrumental in determining device operation and which has furthermore been taken advantage of to image particularly thin ferromagnetic films down to just a few monolayers~\cite{thinSEMPA} and even layered antiferromagnets due to the strong contribution of the uppermost atomic layer~\cite{AFMLSMO}. 

Virtually all SEMPA imaging to date has been static or quasi-static in nature. The long acquisition times are a barrier to investigation of magnetization dynamics, however very recently the feasibility of imaging on nanosecond timescales with advanced signal processing based on the time of detection of the individual electron counts has been demonstrated~\cite{timeSEMPA, timeSEMPA2}. The high spatial resolution is a key advantage of the technique, enabling imaging not only of domain configurations but also of domain wall spin structures~\cite{Pascal}. The impetus to increase the spin signal means that it is usual to operate the SEM at large beam-currents and low voltages of typically 1-3\,kV for which the emission of the spin-polarized low energy secondaries is increased~\cite{polen}. Under these conditions typical resolutions are around 20\,nm, however resolution of better than 5\,nm has been demonstrated~\cite{SEMPA1}.

\section{Scanning Probe Microscopy}

The increasingly low dimensions of magnetic nanostructures and accordingly the size of the magnetic structures, e.g. skyrmions, vortices and domain configurations in general, currently reaching the range of 10\,nm~\cite{Hopster2004}, need advanced high spatial resolution microscopy techniques. The scanning probe methods, i.e. spin-polarized scanning tunneling microscopy (SP-STM) and magnetic force microscopy (MFM), may have lateral resolutions down to the atomic dimensions, providing a considerable advantage over many imaging methods. However, the very high lateral resolution comes at a cost because sample environment and tip preparation causes considerable additional experimental effort. We begin with an overview of different scanning probe techniques and then discuss spin-polarized scanning tunneling microscopy and magnetic force microscopy as the two most common methods in more detail in the separate sections.

\index{spin-polarized scanning tunneling microscopy}
Spin-polarized scanning tunneling microscopy goes down to the utmost lateral resolution of scanning tunneling microscopy, being able to even resolve electronic orbitals smaller than atomic distances~\cite{Tersoff1983,Tersoff1985}. The magnetic contrast is introduced by using a ferromagnetic or antiferromagnetic tip, exploiting the spin-dependent differences of the density of states in tip and sample.
First results of spin-polarized scanning tunneling microscopy were already shown in the early nineties~\cite{Wiesendanger1990}, but it took some years until it became an established method~\cite{Bode1998,Wulfhekel1999}. The difference in tunneling conductivity is in principle similar to the effect exploited in a magnetic tunneling magnetoresistance device. The more states that are available to tunnel, the higher the resulting tunneling current. The tunneling current from the tip is spin-polarized because of the imbalance of electrons with spin-up and spin-down. The current is also proportional to the number of free states that are available for the electrons to tunnel into. Consequently, the current is different for parallel and antiparallel orientations of tip and sample magnetization. Considering atomic resolution, the method can also be applied if one or both surfaces of tip and sample are antiferromagnets avoiding the problem of tip - sample interaction. Spin-polarized scanning tunneling microscopy definitely requires UHV conditions and ultra-clean surfaces. Although results obtained at room temperature have been reported, low temperature experiments considerably increase the mechanical and electronic stability of the measurement.
\\
\\
\index{magnetic force microscopy}
Magnetic force microscopy is in principle very well suited to image magnetic domains with high resolution in an ambient environment. However it is intrinsically limited in lateral resolution by the physical effect used to obtain magnetic contrast. Magnetic force microscopy relies on the long range magnetic dipolar interaction of a magnetic tip and the stray field of the sample. The magnetic tip has to be lifted a few nm above the surface to avoid van-der-Waals interactions, thus decreasing the attainable resolution, while the measurable magnetic force rapidly decreases with distance. This trade-off between resolution and signal limits the obtainable resolution and atomic resolution cannot be achieved~\cite{Wulfhekel1999}. Another drawback is the fact that the tip - sample interaction may easily change the magnetic structure during  scanning. A considerable improvement has been achieved by exploiting the extremely short-ranged magnetic exchange interaction, as proposed and experimentally shown for a prototypical antiferomagnetic material NiO~\cite{Kaiser2007}. \index{atomic force microscopy}By using atomic force microscopy with a magnetic tip one detects the short-range magnetic exchange force between tip and sample spins, revealing the arrangement of both surface atoms and their spins simultaneously. With this technique the inter-spin interactions can be investigated at the atomic level. Since the exchange interaction is strongly modulated by any material between tip and sample, this method only works for very clean surfaces in UHV.

Instead of measuring the force due to the dipolar interaction one can alternatively measure the stray field directly using a Hall probe~\cite{Shono2000,Chang1992, Howells1999} or a superconducting quantum interference device (SQUID)~\cite{Vu1993}\index{Hall probe}\index{SQUID}. Both methods are passive measurements avoiding magnetic perturbation of the specimen. This advantage comes at the cost of less spatial resolution ($\approx 0.35$\,$\mu$m) as determined by the dimensions of the lithographically fabricated probe. Scanning Hall probe microscopy has the advantage of a wider operating temperature range and a decent field sensitivity of 0.1\,G. The SQUID probe operated at low temperatures on the other hand is considerably more sensitive to small fields ($10^{-6}$\,G). A spatial resolution of 10\,$\mu$m has been demonstrated~\cite{Kirtley1995}.

A very special type of magnetic force microscopy is given by magnetic resonance force microscopy (MRFM)\index{magnetic resonance force microscopy}. In this experiment a force signal is generated by modulating the sample magnetization with standard magnetic resonance techniques~\cite{Sidles1995}. The magnetic tip at the end of a cantilever is positioned roughly 100\,nm above the sample surface. The isosurface of constant stray field of the tip defines a resonant slice representing those points in the sample where the field matches the condition for magnetic resonance. As the cantilever vibrates, the resonant slice swings back and forth through the sample causing cyclic adiabatic inversion of the spin. The cyclic spin inversion causes a slight shift of the cantilever frequency owing to the magnetic force exerted by the spin on the tip. Spins as deep as 100\,nm below the sample surface can be probed. By moving the tip in all three dimensions a tomographic image of the spin distribution can be mapped. The main advantage is the outstanding sensitivity of this method, providing single electron spin detection in combination with high spatial resolution of 10\,nm~\cite{Rugar2004}. The sensitivity is large enough to sense nuclear spins, too. Measuring the nuclear spin-lattice relaxation times locally in the mK temperature range allows the characterization of magnetic properties of inhomogeneous electron systems realized in oxide interfaces, topological insulators, and other strongly correlated electron systems such as high-T-c superconductors~\cite{Wagenaar2016}.

Finally nitrogen-vacancy centre magnetometry is a very promising emerging technique which has also been  incorporated into atomic-force microscopes to provide particularly sensitive, high-spatial resolution magnetic imaging~\cite{Rondin2014}. The approach is based on the proposal by Chernobrod and Berman to use single electronic spins as local magnetic field sensors~\cite{Chernobrod}. As the spin is scanned over the surface, the local magnetic field causes a Zeeman splitting of the electronic energy spin sublevels which can be detected optically by measuring the photoluminescence of the probe in an electron spin resonance measurement. The probe spin system of choice is a single nitrogen-vacancy defect in diamond which exhibits the required properties for the measurement including favourably long coherence times. The nitrogen-vacancy defects are created near the surface of diamond nanocrystals or nanopillars via high energy electron/ proton irradiation, followed by annealing. The diamond is then mounted into an AFM to act as the probe-tip. As the tip is scanned across the surface, microwave fields are applied to stimulate electronic transitions between the spin triplet sublevels of the system. On resonance, the photoluminescence spectra show a characteristic drop in intensity. Due to the Zeeman effect, this feature is split and shifted in an applied magnetic field, providing a measurement of the projection of the field along the nitrogen-defect quantization axis that is localized at the defect site. In this manner, the stray magnetic field from a magnetic vortex core~\cite{vcore} and even a single electron spin have been imaged~\cite{espin}. In addition to this high sensitivity, for the imaging of spin textures it can provide an excellent spatial resolution of typically a few 10s of nm, depending on the sample surface to nitrogen-vacancy defect separation. It is also non-invasive and has succesfully been employed to image the pinning and propagation of magnetic domain walls in nanowires~\cite{Tetienne2014} without undesired perturbations of the magnetic state from the tip.

\subsection{Spin-polarized Scanning Tunneling Microscopy}
\label{SPSTM}
\index{scanning tunneling microscopy}
In scanning tunneling microscopy (STM), the apex of a conductive tip is placed near the surface of a conductive sample. A bias voltage is applied between sample and tip and a small tunneling current flows that decays exponentially with the tip-sample separation. In the constant current mode a feed-back mechanism adjusts the tip-sample distance such that the tunneling current is kept constant. When the tip is scanned over the surface, the tip apex moves on lines of constant current, which are related to first order to lines of constant density of states, i.e. reflecting the sample topography. For spin-polarized STM a spin-polarized tunneling current is needed that in principle can be obtained in various ways~\cite{Bode2003,Wiesendanger2009}. 

Before the already mentioned use of ferromagnetic tips is discussed, we shortly report on alternative approaches that have been tried with less success. The possibility to photo-excite spin-polarized carriers from GaAs tips has been considered by Suzuki et al.~\cite{Suzuki1997}. Circularly polarized light was used to pump spin-polarized carriers into the conduction band of the tip that then tunnel into the sample. By modulating the helicity, the tunneling current is modulated due to spin dependent tunneling, which can be detected by a lock-in amplifier. The signal can be used to separate spin information from the sample topography.  However, this method suffers from low contrast and additional magneto-optical contrast of low resolution. As a new development, photoemission or photoassisted tunneling from metallic tips has been proposed for the introduction of time resolution to STM on a femtosecond level~\cite{Terada2010}. Using the inverse effect, Alvarado et al.~\cite{Alvarado1992} measured the circular polarization of the light that is emitted because of the tunneling current. This method suffers from the very low quantum efficiency of the inverse photoemission effect and details of the emission process are still debated today~\cite{Berndt2015}.

For ferromagnetic tips the separation of topography and spin information is an important issue. When a finite negative (positive) bias voltage, $V$, is applied to the sample with respect to the tip, the occupied sample (tip) states in the range of width $eV$ below the Fermi level of the sample (tip) contribute to the tunneling. In the tunneling process, the electrons tunnel into the unoccupied tip (sample) states of the range $eV$ above the Fermi level. The spin polarization of both the tip and the sample states contribute to the tunneling. Therefore, the spin polarization of the tunneling current varies with sample bias. Variations of the tunneling conductance are compensated by the feed-back loop and show up in the topography that then contains both topographic and magnetic information.

An effective way of separating magnetic and topographic information has been demonstrated by Wulfhekel et al.~\cite{Wulfhekel1999}, who modulated the tip magnetization by a small coil and detected the modulation of the tunneling current by a lock-in amplifier. For out-of-plane sensitivity the coil is wound around the tip axis. For in-plane sensitivity one uses a small ring from a soft magnetic material, where the outer rim is used as a tip with surprisingly good resolution. The tip material has to be chosen carefully as magnetostriction produces an additional signal with the same frequency as the modulation.
Domain wall widths in Mn/Fe bilayers in the range of 1\,nm and step-induced frustration of antiferromagnetic order have been resolved in this manner~\cite{Schlickum2004}. The most convenient way of separating magnetic and topographic signals is by comparing two measurements obtained for opposite magnetization directions of tip or sample~\cite{Prokop2005}. One option is to repeat the measurement after an external field has rotated tip or sample magnetization~\cite{Oka2010}. A second possibility is to compare sample areas of supposed identical chemical structure but opposite magnetization, e.g. an epitaxially grown nanowire of constant thickness~\cite{Pratzer2001}. A third option is to repeat the measurement with intentionally changed tip magnetization, e.g. from out-of-plane to in-plane magnetization~\cite{Prokop2006}.

\subsubsection{Technical Details}
\index{scanning tips}
Chemically etched tungsten tips are the most commonly used tips for STM. Starting from these tungsten tips a thin ferromagnetic film evaporated on the tip apex then serves as a ferromagnetic counter electrode. The stray field of the thin film is small enough to avoid dipolar sample-tip interaction. Alternatively, antiferromagnetic tips are used. In this case Cr tips etched from thin Cr wires are advantageous. For SP-STM the tip must be prepared in-situ in UHV, for example by voltage pulses, in order to obtain high resolution and magnetic contrast. Usually, only a single magnetization component is detected. The sensitivity axis of the tip is in many cases not obvious and must be calibrated on known magnetization structures.

The tunneling current can be described as a sum of a spin-averaged and a spin-dependent term~\cite{Wortmann2001}. Following Ref.~\cite{Bardeen1961}, the tunneling current can be calculated by Fermi's golden rule:
\begin{equation}
I\propto \int_{\infty}^{\infty}|M(\varepsilon)|^2\rho_T(\varepsilon-eV)\rho_S(\varepsilon)[f(\varepsilon-eV)-f(\varepsilon)]d\varepsilon.
\end{equation}
Here $M$ denotes the tunneling matrix element, $\rho_{T,S}$ is the spin-dependent density of states for tip and sample, respectively, and $f$ is the Fermi function. As a result of the integration, the spin-polarized contribution to the current becomes reduced if the spin polarization changes sign between $\varepsilon_F$ and $eV$. In order to increase the magnetic contrast and also for separation of topographic and magnetic information it is helpful to measure the differential conductance $dI/dV$, which in the case of low temperature and low bias is given by~\cite{Bode2003}:
\begin{equation}
dI/dV(V) \propto |M(\varepsilon)|^2\rho_T\rho_S(eV-\varepsilon_F).
\end{equation}
Here we have assumed an energy independent value for $\rho_T(\varepsilon_F)$. By introducing the spin polarization for tip, $P_T(\varepsilon_F)$, and sample $P_S(eV-\varepsilon_F)$ the differential conductivity can be written as 
\begin{equation}
dI/dV(V) =g_0(V)[1+P_T(\varepsilon_F)P_S(eV-\varepsilon_F)cos\phi],
\end{equation}
with $\phi$ denoting the angle between tip and sample magnetization and $g_0$ the unpolarized conductivity. The assumption of a constant tip polarization $\rho_T(\varepsilon_F)$ is realistic in the case of $V>0$ probing the unoccupied states of the sample because the tunneling current is dominated by electrons from the Fermi level of the tip. For illustration of this case the tip and sample density of states is sketched on a common energy scale in Fig.~\ref{fig:Methfessel2011}(a). In contrast, when probing the occupied sample states $V<0$ a strong convolution with the tip density of states has to be considered.

\subsubsection{Experimental Examples}

\begin{figure}[t]
\includegraphics[width=\columnwidth]{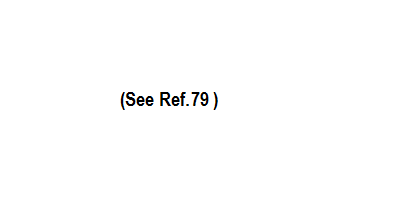}
\caption{\label{fig:Methfessel2011}
(a) Sketch of the tip and sample density of states in a common energy scale assuming positive sample bias.
(b) Topographical image of a Cu-phthalocyanin molecule on Fe(110).
$dI/dV(V)$-spectra (c,g,e) and asymmetries (d,f,h) extracted from the indicated areas in (b) on the Cu-phthalocyanin molecule deposited on Fe(110).
$dI/dV(V)$-spectra are measured for parallel (black) and antiparallel (red) orientation of tip and sample magnetization.
Experimental asymmetries (blue) are compared to calculated asymmetries (black lines). For comparison, the asymmetry of clean Fe
(d) scaled by 0.5, are also shown in red lines in (f) and (h). 
(See Ref.~\cite{Methfessel2011} T. Methfessel, S. Steil, N. Baadji, N. Grossmann, K. Koffler, S. Sanvito, M.
Aeschlimann, M. Cinchetti, H. J. Elmers, Phys. Rev. B 84 224403 (2011).)
}
\end{figure}

The imaging of molecular structures is an important step towards the understanding of  spin transport and scattering in hybrid organic-metallic interfaces~\cite{Schwoebel2012}. \index{organic-metallic interfaces}
An example of a Cu-phthalocyanin molecule on a metallic ferromagnetic surface is shown in Figure~\ref{fig:Methfessel2011}~\cite{Methfessel2011}. Figure~\ref{fig:Methfessel2011}(c-h) shows $dI/dV$ spectra and the corresponding spin asymmetry, defined as $A=[D(\uparrow\uparrow)-D(\uparrow\downarrow)]/[D(\uparrow\uparrow)+D(\uparrow\downarrow)]$. From the spectroscopic results obtained by collecting the current at tip positions over the molecule one can observe the almost entire suppression of the peaks at $-0.1$\,eV, as compared to the clean Fe surface. In contrast a new peak appears at $+0.4$\,eV which is attributed to electronic states that originate from the hybridization of the lowest unoccupied molecular orbital (LUMO) of the free molecule with the substrate. By comparing the asymmetry of the clean Fe layer to that of the CuPc molecules two regions are distinguished, denoted as regions 1 and 2. In region 1 the spectral features of the asymmetry are only little modified with the exception of a global reduction of $A$ by about 50\%, which is explained by assuming that the Fe-CuPc interface acts as a featureless scattering barrier. The pronounced deviations in dependence on position in the two regions 2 is explained by the presence of spin-polarized hybridized interface states.

The observation of atomic scale magnetic skyrmions in ultrathin magnetic films also highlights the ultra-high resolution of spin-polarized STM for magnetic microscopy~\cite{Heinze2011}. The nontrivial spin textures are topologically stable, particle-like spin configurations that can be used as information carriers. Spin polarized scanning tunneling microscopy can not only be used for the imaging of skyrmions
but also for writing and deleting individual skyrmions~\cite{Romming2013} by employing the tunneling electrons.

\subsection{Magnetic Force Microscopy}
\label{MFM}
\index{magnetic force microscopy}
MFM is a special operation mode of atomic force microscopy employing a magnetic probe, which interacts with the magnetic stray fields of the sample~\cite{Martin1987}. 
Therefore, this technique measures the stray field distribution rather than the magnetization structure itself.
Recent developments are focussed on the quantitative analysis of data, improvement of resolution, and the application of external fields during measurement~\cite{Schwarz2008}.
The interpretation of images acquired by MFM requires knowledge about the specific near field magnetostatic interaction between probe and sample. In addition one has to consider the properties of suitable probes. More details can be found in Refs. \cite{Schwarz2008,Hartmann1999,Rugar1990,Hug1998}.

\subsubsection{Technical Details}

For the measurement of the magnetic forces almost exclusively the dynamic mode is applied, where resonance frequency shifts of the oscillating cantilever are measured either directly or indirectly by the amplitude variation for fixed excitation frequency. The frequency of the oscillating is given by 
\begin{equation}
f = f_0\sqrt{1-\frac{1}{c^*}\frac{dF_m}{dz}}
\end{equation}
with $f_0$ being the resonance frequency without interaction, $\frac{dF_m}{dz}$ the gradient of the magnetic force and $c^*$ an experimental constant. The sign of the frequency shift distinguishes between attractive and repulsive forces ($\Delta f < 0$ and $\Delta f > 0$, respectively). The most common detection method uses the amplitude signal and is referred to as amplitude modulation. The cantilever is driven slightly away from resonance, where the slope of the amplitude-versus-frequency curve is large. Measurement sensitivity has an inverse dependence on the $Q$ value of the oscillating system. However a high $Q$ value has the drawback of an increased response time of the detection system. In this case a suitable alternative is the frequency modulation (FM) technique. The cantilever self-oscillates with constant amplitude $A_0$, with tip-sample interactions shifting the actual cantilever frequency $f$ by $\Delta f  =  f - f_0$. 

The standard MFM probes are etched silicon tips with magnetic coatings consisting of 10-150\,nm Co/Cr multilayer structures and an effective magnetic moment of around $10^{-22}$\,Vsm. However, a large variation of materials have been applied. A large coercive field is favorable in order to avoid a change of the magnetic configuration of the tip during scanning.

For the separation of topography and magnetic signal a constant distance mode is applied. This lift mode involves measuring the topography on each scan line in a first scan and the magnetic information in a second scan of the same line. This height data of the first scan is used to move the tip at a constant local distance above the surface during the second (magnetic) scan line, during which the feedback is turned off.
At this larger distance the topographic interaction has decreased to a level that it does not overlay the magnetic interaction, which decreases with distance at a smaller rate.

\subsubsection{Experimental Examples}

As an example of high resolution imaging using MFM the technique has been employed to image of bit patterned media with perpendicular anisotropy where a resolution of better than 10\,nm has been demonstrated by evaluating line profiles in the images~\cite{Belova2012}.

\begin{figure}[b!]
\centering
\includegraphics[width=0.8\columnwidth]{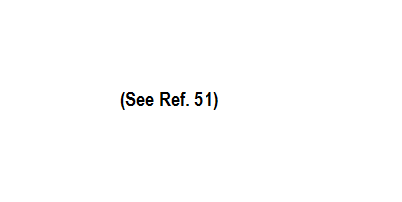}
\caption{\label{fig:Kaiser2007}
MFM images using the same tip at 7.9\,K in 5\,T on the same area. 
(a) Image recorded at a constant frequency shift of -22.0\,Hz. 
(b) Image measured at a constant frequency shift of -23.4\,Hz corresponding to a reduction of tip - sample distance of 30\,pm.
(c,d) Fourier transforms of (a) and (b), respectively.
(See Ref.~\cite{Kaiser2007}) Nature U. Kaiser, A. Schwarz, R. Wiesendanger,  446, 7135,  (2007). }
\end{figure} 

A considerable increase in spatial resolution can be achieved by magnetic exchange force microscopy. The general concept of magnetic exchange force microscopy relies on the combination of the atomic resolution atomic force microscopy with spin sensitivity by using as a force sensor a magnetic tip mounted on the free end of a cantilever. During scanning in the $x–y$ plane, $\Delta f$ is kept constant by adjusting the $z$ position of the tip relative to the surface so that the recorded topographic image represents the condition of a constant tip - sample interaction force. Selecting a more negative $\Delta f$ set-point increases the attractive interaction; that is, the tip - sample distance is reduced. This method permits atomic resolution on conducting and non-conducting surfaces in the non-contact regime with height differences (or contrast) 
in the topography image reflecting variations of the short-range forces. A purely chemical and structural contrast would reflect only the arrangement of atoms. If a magnetic exchange interaction between tip and sample is present, an additional contrast modulation occurs between neighboring rows of magnetic atoms in an otherwise identical chemical environment. For this reason, the exchange interaction can be distinguished unambiguously from other tip - sample interactions. For the illustration of the method we show a result of Kaiser et al.~\cite{Kaiser2007} obtained for the surface of the antiferromagnet NiO.

Fig.~\ref{fig:Kaiser2007} shows two atomically resolved images for NiO(001). Both images were acquired on the same sample area. The topographic image Fig.~\ref{fig:Kaiser2007}(a) recorded at a smaller frequency shift exhibits the $(1\times 1)$ symmetry of the chemical surface unit cell. In the Fourier transform of the data the chemical unit cell is represented by four spots. Fig.~\ref{fig:Kaiser2007}(b) acquired at a larger frequency shift, i.e. at smaller sample tip distance, shows an additional modulation: every second row of nickel atoms along the [110] direction seems more depressed, as indicated by the black arrows. 
The corresponding Fourier transform [Fig.~\ref{fig:Kaiser2007}(d)] exhibits the appearance of one additional pair of peaks located halfway between the center and two (opposing) peaks corresponding to the chemical unit cell. This additional contrast modulation on neighboring nickel rows reflects the antiferromagnetic surface unit cell of NiO(001).

\section{X-ray Imaging}
~\index{x-ray imaging}
\label{sec:xray}
In this section we review the working principle as well as the strengths and weaknesses of the most established x-ray imaging techniques. Specifically, we discuss transmission x-ray microscopy (TXM), scanning transmission x-ray microscopy (STXM), photoemission electron microscopy (PEEM), and coherent diffractive imaging (CDI). In the end, we also briefly present a technique for band structure or momentum space imaging of magnetic materials, namely spin-polarized angle resolved photoemission spectroscopy (SP-ARPES). Reviews of imaging with x-ray microscopy can be found in refs.~\cite{xray1, xray2}.

\subsection{X-ray Magnetic Circular Dichrosim -- a contrast mechanism}
\label{subsec:XMCD}
~\index{x-ray magnetic circular dichroism}
X-ray magnetic circular dichrosim (XMCD) is the contrast mechanism for all real-space x-ray magnetic imaging techniques~\cite{schutz}. The XMCD effect describes how the absorption of photons at a specific energy depends on the relative orientation of the local magnetization and the helicity of the photons. Here, we briefly review the excellent text on XMCD by St\"ohr and Siegmann \cite{stohr_magnetism_2006}.

\begin{figure}[t]
\centering
\includegraphics[width=0.8\textwidth]{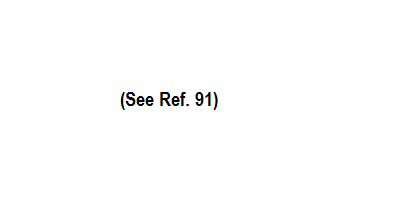}
\caption{Left: Schematic illustration of the XMCD effect. X-ray illumination of the sample excites element specific electronic transitions from core-levels to empty states at the Fermi level. Due to the spin-split density of states in the magnetic specimen, the x-ray absorption cross-sections are different for negative and positive helicity of the incident photons, depending on the relative alignment of the incident wave-vector and the magnetization in the sample, as shown on the right in the case of the L$_2$ and L$_3$ absorption edges of Fe. (See Ref. \cite{stohr_magnetism_2006}, J. St\"{o}hr and H. C. Siegmann, Magnetism - From Fundamentals to Nanoscale Dynamics (2016)}
\label{fig:XMCD}
\end{figure}

Consider the L-edge resonant photon absorption of a $3d$ magnetic transition metal, i.e., the excitation from the localized $2p$ level to the delocalized and spin-polarized $3d$ band, as depicted in Fig.~\ref{fig:XMCD}. This transition is well described as a first order dipole transition. The selection rules for such dipole interactions require $\Delta \ell = \pm 1$, $\Delta s = 0$, $\Delta m_l = q$, and $\Delta m_s = 0$, where $\ell$ is the magnitude of the orbital angular momentum, $m_l$ is the magnetic quantum number corresponding to the orbital angular momentum, $s=\frac{1}{2}$ is the spin angular momentum, $m_s$ is the spin orientation, and $q$ is the orbital angular momentum of the incident photon, which is $1$ for right circularly polarized photons and $-1$ for left circularly polarized photons (all in units of $\hbar$). In particular, we will use the fact that $m_l$ changes by $q$ and that $m_s$ is conserved.

The $2p$ levels experience a strong spin orbit coupling, typically on the order of \SI{15}{eV}, which ensures that the L$_2= 2p_{1/2} \rightarrow 3d$ and the L$_3= 2p_{3/2} \rightarrow 3d$ absorption edges do not overlap. Consider the L$_2$ transition. The total angular momentum of the initial $2p_{1/2}$ state is $j=1/2$ and the possible values of $m_j = m_l + m_s$ are $\pm 1/2$. Hence, for up spins ($m_s=1/2$), the two possible values for $m_l$ are $m_l=0$ ($m_j=1/2$) and $m_l=-1$ ($m_j=-1/2$). Similarly, for down spins, $m_l$ can take values of $1$ and $0$. All of these states are present with equal probability. If the sample is exposed with right circular light ($q=1$), the selection rules require that the $m_l$ quantum number is increased by $1$. That means that up spins are excited to $m_l=0$ and $m_l=1$ whereas down spins are excited to $m_l=1$ and $m_l=2$. Spin-orbit coupling in the 3$d$ band is negligible and all $m_l$ values are present with equal probability. However, the transition probability for $m_l=1\rightarrow 2$ is lower than for $m_l=-1\rightarrow 0$, as described by the Clebsch-Gordon coefficients, and the transition $m_l=0\rightarrow 1$ is present in both cases. Therefore, the total transition matrix element is smaller for down spins than for up spins and hence the electrons excited at the L$_2$ edge by right circular light have a net spin up polarization. Left-circular light excites electrons with the same spin polarization in the down direction.

The total absorption cross section is proportional to the density of initial states, the excitation probability, and the density of available final states. The first two factors are identical for both helicities. Therefore, the absorption of right circular light minus the absorption of left circular light measures the difference of available up states minus the density of available down states in the $3d$ band, which is proportional to the magnetization of the sample along the photon propagation direction. Hence, the absorption cross section of circular light depends on the relative orientation of the sample magnetization and the photon helicity. 

Imaging with XMCD contrast requires highly monochromatic circularly polarized x-rays at the energies of the L edge of the magnetic transition metals, typically around \SI{800}{eV}, also called soft x-rays. Such light is available with high intensity at modern synchrotrons and free electron lasers and up to now most x-ray magnetic imaging is performed at these facilities. However, the development of high harmonic generation sources has made tremendous progress recently and it seems likely that lab-based soft x-ray imaging can become practical in the near future \cite{fan_bright_2015,willems_probing_2015}.

\subsection{TXM -- quick full-field imaging in transmission geometry}
\label{subsec:TXM}
~\index{transmission x-ray microscopy}
Transmission x-ray microscopy (TXM)\cite{fischer_magnetic_2013,sakdinawat_nanoscale_2010} can be seen as an analog to visible light microscopy, with enhanced resolution by using smaller probing wave lengths. Lenses at the wavelengths of soft x-rays are realized by diffractive elements, so-called zone plates. The far field diffraction pattern of a specimen is given by the Fourier transform of its transmission function. A focus, i.e., a point-like diffraction pattern, can be obtained from a Bessel function transmission function. A zone plate is a binary version of a Bessel function absorption mask. Ultimately, the focus size of a zone plate is determined by the width of the outermost zone. High resolution zone plates are difficult to fabricate and therefore very expensive.

\begin{figure}[t]
\centering
\begin{minipage}[t]{95mm}
\centering
\vspace{0pt}
\begin{overpic}[width=\hsize]{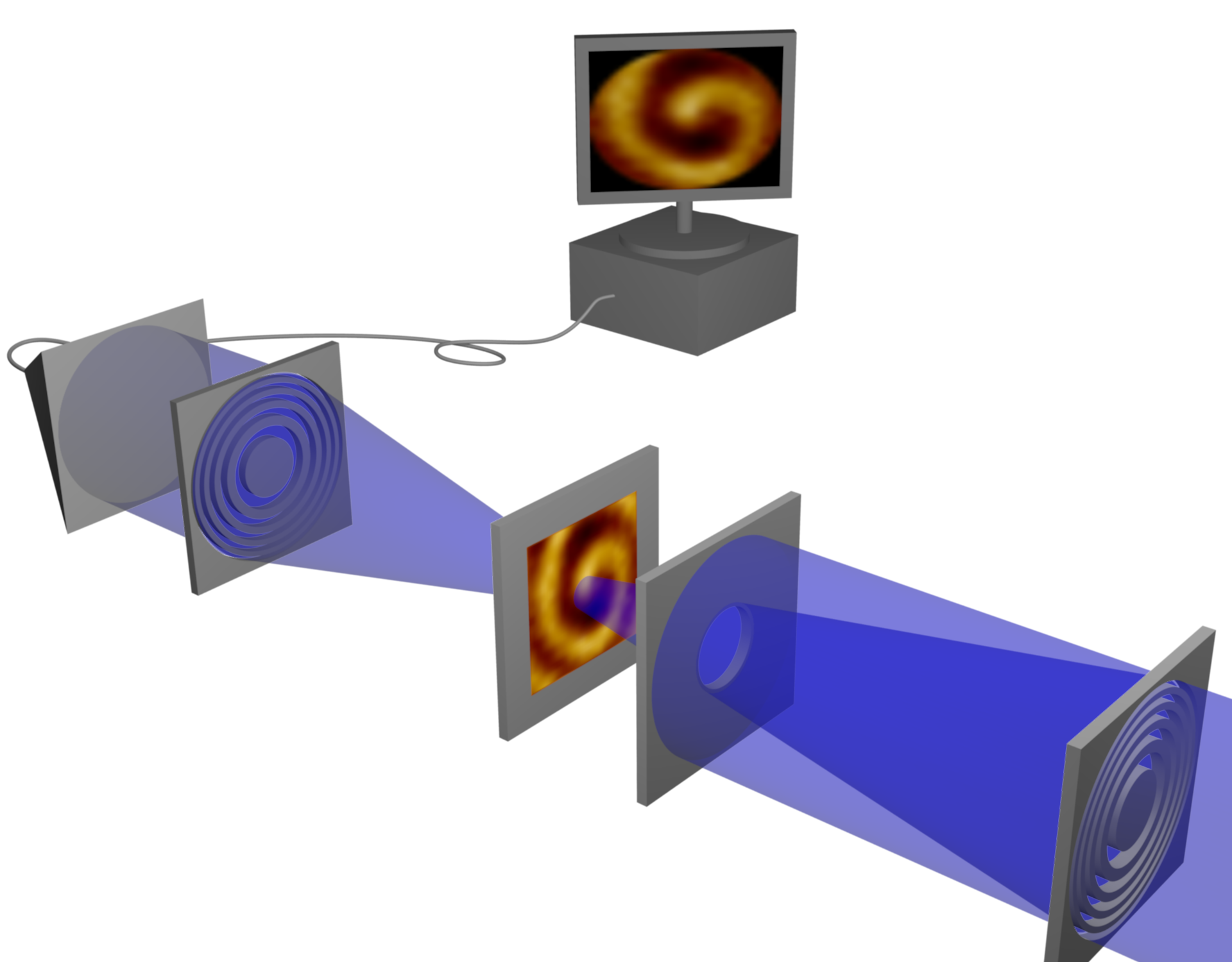}
\put(63,39){\textbf{OSA}}
\put(93,28){\textbf{KZP}}
\put(47,43){\textbf{Sample}}
\put(26,53){\textbf{MZP}}
\put(10,55){\textbf{CCD}}
\end{overpic}
\end{minipage}
\caption{Schematic illustration of a transmission x-ray microscope. The incident beam is transmitted through a condenser zone plate (KZP), which is made of alternating opaque and transparent rings to mimic a Bessel transmission function. The non-diffracted zero order light, as well as higher order diffractions, are blocked using an order selecting aperture (OSA). The sample is placed close to the focus of the KZP. An image of the transmitted light is generated via an objective zone plate (MZP) on a CCD camera chip}
\label{fig:TXM_schematics}
\end{figure}

The concept of TXM is illustrated in Fig.~\ref{fig:TXM_schematics}. Similar to an optical microscope, TXM employs a condenser (KZP) that reduces the spot size of the incoming light to the field of view of the subsequent objective lens, i.e., to a circle of approximately \SI{10}{\micro m} in diameter. Like every zone plate, the condenser has a limited efficiency on the order of \SI{10}{\percent}. The majority of the transmitted light is undiffracted zero order light, which is blocked by an order selecting aperture (OSA). The focus of the condenser depends on the wave length of the incoming light. The position of the OSA is optimized for transmitting the first order light of the required wavelength, blocking all other wavelengths because of their different cone angles. Hence, the OSA also acts as a monochromator.

The light transmitted through the sample is collected by an objective zone plate lens (MZP) and transformed to a real-space image of the local transmission intensity of the sample on a CCD camera. Typically, the CCD camera has 2048 pixels per line and is operated in $2\times 2$ binning mode, resulting in \SI{10}{nm} pixel size for the \SI{10}{\micro m} field of view. The binning allows for low noise readout in \SI{1}{s} and a good quality image is obtained by accumulating 20 images per helicity, yielding a full XMCD image in approximately \SI{1}{min}.

The fast acquisition of large scale images with good resolution is the major advantage of TXM compared to other magnetic imaging techniques. Furthermore, some TXM end stations are built such that the OSA and the MZP act as vacuum windows and the sample is in air, which is helpful for some applications. However, due to the full-field nature of the technique, fast multi-pixel detection is needed for dynamic imaging, limiting the repetition rate for dynamic processes.

\subsection{STXM -- optimized for dynamic imaging}
\label{subsec:STXM}
 ~\index{scanning transmission x-ray microscopy}
Scanning transmission x-ray microscopy (STXM) \cite{sakdinawat_nanoscale_2010, kilcoyne} is similar to TXM, but instead of collecting a full field image with an objective zone plate the sample is scanned with high precision through a focussed x-ray spot and the total transmission is detected by a fast avalanche photo diode (APD). The resolution is now determined by the spot size of the incident photons, which is typically \SI{25}{nm} but can be significantly smaller with more sophisticated zone plates (usually at the loss of total intensity) or using ptychography~\cite{ptychography}. The readout of the APD is extremely fast, indeed faster than the temporal separation, $\delta t$, of subsequent x-ray flashes (e.g., \SI{2}{ns} in the multi bunch mode at the BESSY II synchrotron). Typically, a configurable number of channels, $n$, is available for counting the transmitted photons. Provided the investigated sample is excited with an excitation of period $\delta t\,n/m$ (with $m,n$ coprime integers, $m$ often called ``magic number''), a movie of the response of the sample is directly obtained from the images collected in the $n$ different channels. The frames of such a movie can be reshuffled in a pulse-chronological order, such that the dynamic response to the excitation is scanned in $n$ temporal steps of $\delta t/m$. Ideally, the number of channels is coprime to the number of bunches in the synchrotron ring, such that the light from every bunch contributes equally to each channel and bunch fluctuations are averaged out.

STXM is a very versatile technique. It allows for almost arbitrary zooming and real-space translation of the field of view. That is, a large number of objects can be investigated without changing samples. The capability of recording movies directly makes the technique favorable for dynamic imaging. Typically, a single XMCD image of a \SI{3}{\micro m} field of view can be obtained in less than 2 minutes and a full movie with hundreds of frames can take less than \SI{30}{min}.

\subsection{PEEM -- imaging surfaces of bulk samples}
\label{subsec:PEEM}
~\index{photoemission electron microscopy}
\begin{figure}[b]
\centering
\begin{minipage}[t]{95mm}
\centering
\vspace{0pt}
\begin{overpic}[width=\hsize]{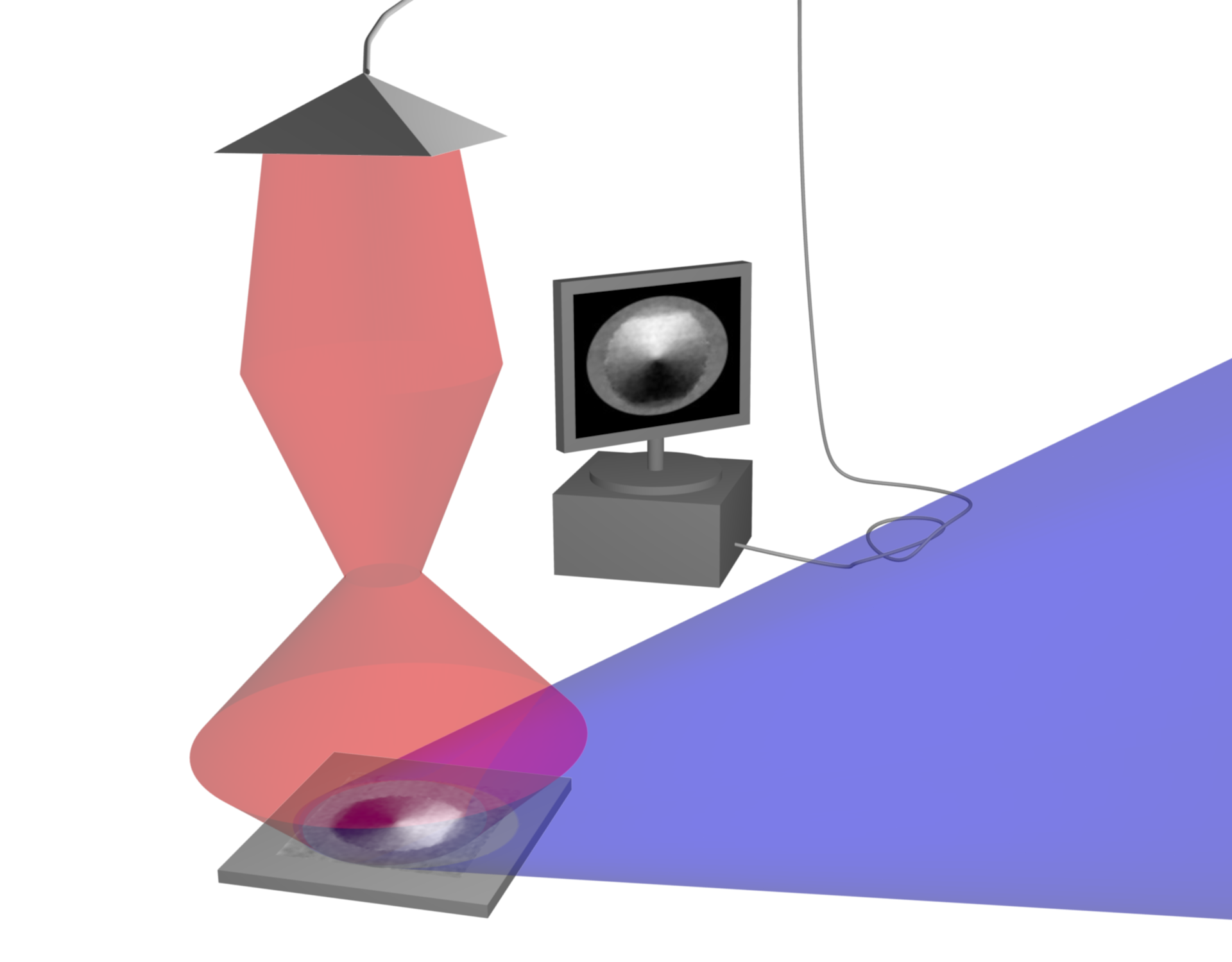}
\put(88,28){\textbf{X-rays}}
\put(-.2,11.5){\SI{20}{kV}}
\put(22,53){\textbf{electrons}}
\put(11,71){\textbf{detector}}
\put(7,10){
\begin{tikzpicture}
\draw[thick] (.9,0) -- (0,0) -- (0,.2) -- (-.1,.2) -- (.1,.2);
\draw[thick] (.5,.5) -- (0,.5) -- (0,.3) -- (-.15,.3) -- (.15,.3);
\end{tikzpicture}}
\end{overpic}
\end{minipage}
\caption{Schematic illustration of PEEM. The incident x-rays hit the sample under grazing incidence, hence being absorbed in the surface near region. The larger the absorption cross section the more energy is deposited near the surface, where secondary electrons are generated that can leave the sample into the vacuum. The free electrons are accelerated by a strong electric field of \SI{20}{kV} towards a detection column, where they are energy filtered and focussed to a 2D detector. For simplicity, the electron optics are not drawn in the schematic}
\label{fig:PEEM}
\end{figure}
Photoemission electron microscopy (PEEM) makes use of the fact that an electron excited by a photon absorption can relax by transferring its energy to other electrons at the Fermi energy~\cite{PEEM1, PEEM2}. If this release of energy to the Fermi level happens near the surface of the sample, some electrons receive enough energy to leave the sample into the vacuum. These free electrons can be accelerated by a high voltage and focussed by a series of electromagnetic lenses to a pixel detector, see Fig.~\ref{fig:PEEM}. The number of electrons at a specific pixel of the detector is proportional to the number of electrons emitted at the corresponding position of the sample, which in turn is proportional to the absorption of photons in the surface-near region. Making use of the XMCD effect, an image of the magnetization in the direction of the incident x-rays can be obtained. Due to the low mean free path of secondary electrons inside the material the sensitivity is restricted to the first few nm below the surface. To increase the absorption efficiency in this surface near region PEEM is typically operated with the x-rays hitting the sample at grazing incidence. PEEM can also be realized in a scanning configuration \cite{sakdinawat_nanoscale_2010}.  Time-resolved measurements are possible by synchronizing an external excitation of the sample with the x-ray imaging pulses, thus allowing for pump-probe type measurements. In contrast to the other x-ray imaging techniques discussed here, PEEM does not require an x-ray transparent sample. Therefore, bulk substrates of arbitrary thickness can be used, which makes PEEM very attractive to investigate epitaxial samples. A constraint is that the sample surface must be conducting. PEEM is also a favorable technique for imaging of the in-plane components of the magnetization due to the grazing incidence geometry.

\begin{figure}[t]
\centering

\includegraphics*[width=.9\textwidth]{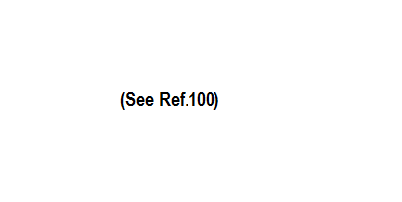}
\caption[]{Dynamic pump-probe PEEM imaging of the field-induced displacement of a transverse domain wall. (a) and (b) show snapshots of the domain wall profile at different time delays, revealing the wall motion. (c) By comparing the displacement field pulse (black) to the wall displacement (green), the delayed onset of the wall motion is observed, indicative of domain wall inertia. Furthermore, by subtracting the running average from the displacement, damped oscillatory motion is revealed (data in blue, fit in red). (See Ref. ~\cite{Rhensius2010}, J. Rhensius, L. Heyne, S. Krzky, L. J. Heyderman, L. Joly, F. Nolting, M. Kl\"{a}ui, Phys. Rev. Lett. 104 067201 (2010) )}
\label{fig:PEEMDW}    
\end{figure}

One example of PEEM imaging is presented in figure~\ref{fig:PEEMDW} where pump-probe techniques have been employed to reveal the dynamic response of a magnetic domain wall to a displacing field pulse~\cite{Rhensius2010}. A transverse domain wall is placed in the centre of a $2$\,$\mu$m wide half-ring by a static external field, as shown in (a). Repeated pulsed in-plane fields are then generated to displace the domain wall by using 15\,ps laser pulses, at a repetition rate of 63\,MHz, which are synchronized with the 70\,ps x-ray imaging pulses from the synchrotron. The laser pulses generate an electrical current pulse via a photodiode and this in-turn is passed through a stripline to generate short, fast-rise time field pulses as shown by the black trace in (c). Pump-probe measurements with varying delay times then provide access to the dynamic motion of the domain wall following the field pulse. A snapshot of the domain wall displacement 200\,ps following the onset of the field pulse is shown in (b) and the full time evolution of the domain wall displacement for a line scan through the centre of the wall is seen in (c). The onset of the wall motion is observed to be delayed with respect to the field pulse and then the wall is observed to undergo damped oscillations (see inset), indicative of domain wall inertia, which can be explained due to observed domain wall spin-structure changes before the motion begins which act as an energy reservoir due to the increase of exchange energy. By fitting the data, a domain wall oscillation frequency of 1.3\,$\pm$\,0.6\,GHz has been extracted and a corresponding domain wall mass of (1.3\,$\pm$\,0.1)$\times 10^{-24}$\,kg deduced.

\subsection{CDI -- zero drift and femtosecond temporal resolution}
\label{subsec:CDI}
~\index{coherent diffractive imaging}
Coherent diffractive imaging (CDI) \cite{chapman_coherent_2010} is a common term for lensless imaging techniques. Generally, the sample is illuminated with a coherent photon beam and the far field scattering pattern, i.e., the squared magnitude of the Fourier transform of the sample's transmission function, is recorded using a camera. The phase information of the scattered wave is lost upon detection and the reconstruction of the transmission function of the sample is non trivial. There are multiple approaches to reconstruct the phase information. For example, known characteristic features of the transmission function, such as an artificial binary mask before the sample, can be used to computationally reconstruct the phase through iterative phase retrieval algorithms. Phase retrieval is particularly popular in hard x-ray imaging and single shot imaging of three dimensional structures \cite{chapman_coherent_2010}. 

\begin{figure}[t]
\centering
\begin{minipage}[t]{95mm}
\centering
\vspace{0pt}
\begin{overpic}[width=\hsize]{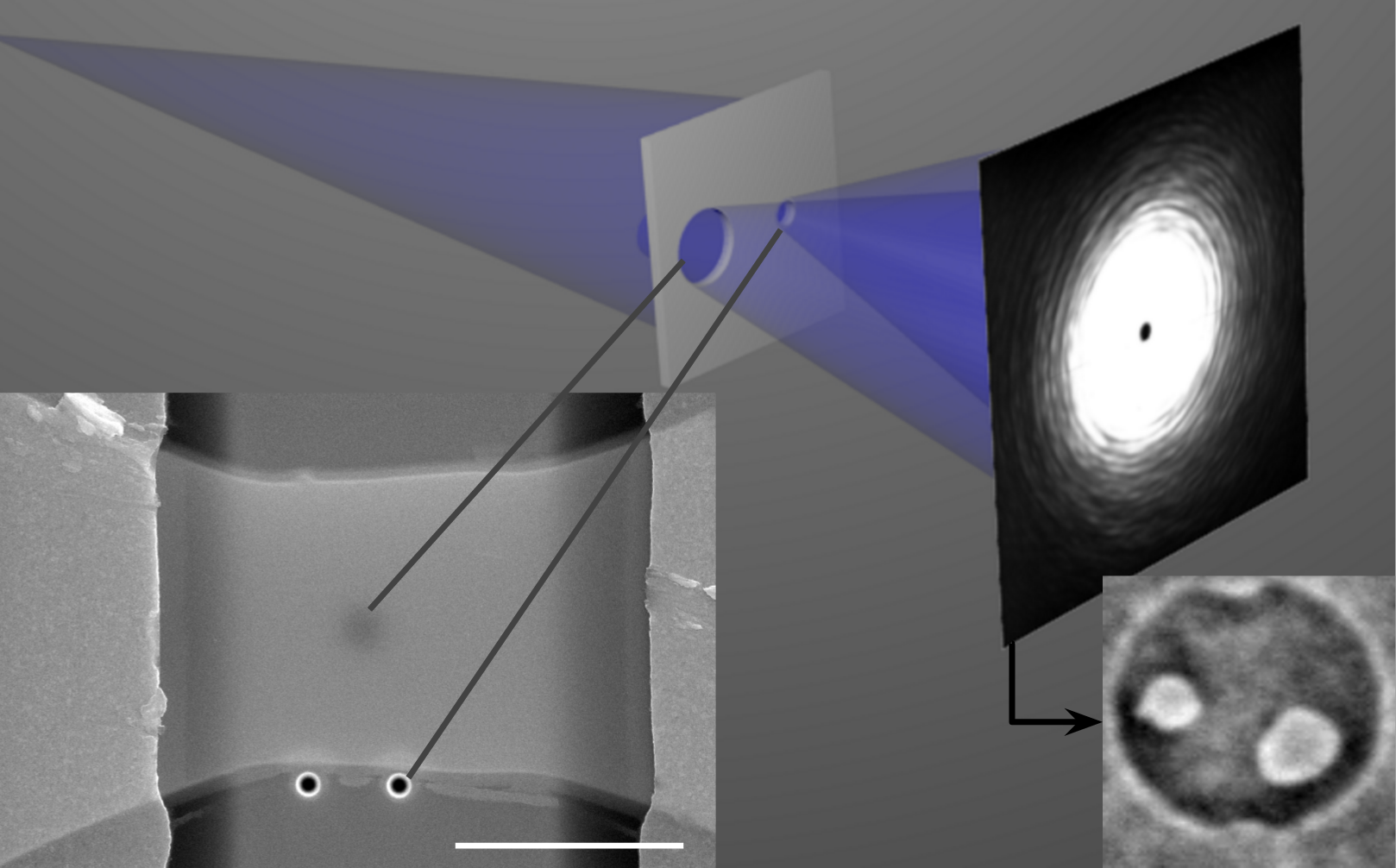}
\put(45,57){\textbf{mask}}
\put(67,57){\textbf{hologram}}
\put(61,11){\textbf{FFT$^{-1}$}}
\end{overpic}
\end{minipage}
\caption{Schematic illustration of x-ray holography. The main image illustrates the principle of x-ray holography, where a coherent x-ray beam (blue) illuminates a specimen behind a circular aperture and a close-by reference hole. The scattered beams of these two objects interfere on a CCD camera chip, forming the hologram. The transmission function of the sample can be reconstructed from the hologram via an inverse Fourier transform. Within the field of view, the reconstruction shows two magnetic skyrmions (white) on a black background. Inset: Scanning electron micrograph of a typical sample. The horizontal wire in the center is a magnetic multilayer. The large features at the left and right edges of the image are gold contact pads for dynamic experiments. The bright vertical rectangle is a SiN membrane. On the back side of the membrane is a \SI{1.5}{\micro m} thick gold layer with three holes in it, one \SI{800}{nm} diameter hole defining the field of view of the imaging (visible as a black shadow in the center of the image) and two smaller holes defining the reference beam (visible as two circles below the magnetic wire). The white scale bar is 5 um long.}
\label{fig:FTH}
\end{figure}

The major drawback of phase retrieval imaging is the highly sophisticated image reconstruction process that is required. Reliable and accurate reconstruction is computationally expensive and requires deep knowledge of the algorithms and their pitfalls. Recently, the concept of ptychography has been adapted to improve the robustness of CDI phase retrieval by recording scattering patterns from multiple largely but not completely overlapping regions of the specimen. However, up to now, phase retrieval and even ptychography is often too time-consuming and involved to be competitive with other soft x-ray techniques when is comes to magnetic imaging~\cite{ptychography}.~\index{ptychography} Ptychography is in this field mainly used for ultra-high resolution imaging. For most applications aiming for \SI{20}{nm} to \SI{50}{nm} spatial resolution, Fourier transform holography (FTH) \cite{eisebitt_lensless_2004, FTH} has become the CDI method of choice. ~\index{Fourier transform holography} The concept is illustrated in Fig.~\ref{fig:FTH}. Essentially, the phase problem is solved by interference of the scattered beam with a reference beam from a point-like reference source. If the reference source is laterally separated from the specimen by a vector, $\vec{r}_0$, then the reference transmission function is given by $\delta(\vec{r}-\vec{r}_0)$. Applying a Fourier transform to the scattering interference pattern, the so-called hologram, yields the autocorrelation of the total transmission function. This autocorrelation includes the cross correlations between the specimen and the reference delta function, hence reconstructing the specimen's transmission function without any sophisticated algorithm. Fig.~\ref{fig:FTH} shows an example of a sample used to image magnetic skyrmions and a reconstruction of the magnetic pattern obtained from an inverse Fourier transform of the hologram.

CDI has its strengths in drift-free imaging and in single shot destructive imaging at free electron laser (FEL) sources~\cite{FEL1, FEL2}. Since the recorded scattering pattern is Fourier space information, drifts of the sample with respect to the detector are translated to a phase shift in the reconstruction and eventually a loss of contrast. However, if the beam aperture and the specimen are rigidly connected, real-space drift of the reconstructed image is fundamentally excluded. It has been demonstrated that moving skyrmions can be tracked with \SI{3}{nm} precision due to this intrinsic stability \cite{buttner_dynamics_2015}. Furthermore, lensless imaging is so far the only viable technique for single shot imaging at FELs, where the beam intensity is so high that all optical elements would be destroyed. However, these advantages come at the price of a sophisticated sample fabrication process and, up to now, a lack of permanent user facility end stations.

\subsection{SP-ARPES -- microscopy in momentum space}
\label{subsec:SPARPES}
~\index{spin-polarized angle resolved photoemission spectroscopy}
Beyond real-space imaging of magnetic domains, the local spin-dependent band structure is of high relevance. Angle-resolved photoemission spectroscopy (ARPES) with spin-detection is capable of providing such information~\cite{ARPES}. ARPES analyzes the electrons released from the sample surface upon photon irradiation. However, instead of secondary electrons (as in the case of PEEM) electrons emitted by direct transitions are detected. In this case energy and momentum conservation apply. Thus, initial state properties, i.e. binding energy, momentum and spin can be deduced from the measured spectra. Recently, a time-of-flight momentum microscope~\cite{elmers2015} has been developed that is capable of parallel detection of momentum, energy, and spin. The electron-optical set-up is related to PEEM but instead of the sample surface the backfocal (or diffraction) plane is imaged on the detector. For more details the reader is referred to the specialized literature \cite{fujiwara_soft_2015, elmers2015, suga_photoelectron_2015,tusche_spin_2015}.

\section{Medical Magnetic Imaging}
\label{sec:medical}
A number of magnetism-based imaging techniques exist which are employed in a medical context~\cite{medical}. In this section we provide an overview of two of the main modern magnetic based techniques, magnetic resonance imaging (MRI) and magnetoencephalography (MEG). 

\subsection{Magnetic Resonance Imaging}
\label{sec:MRI}
~\index{magnetic resonance imaging}~\index{magnetic resonance imaging}
Magnetic resonance imaging employs the phenomenon of nuclear magnetic resonance (NMR) in order to detect the high-field induced polarization of the nuclear spin state of selected species. Most commonly the technique is set to be sensitive to protons which are in high abundance in water and fat rich regions of the body. Depending on the precise sequence of magnetic field pulses applied during the measurement, the technique provides a wide range of contrast mechanisms, allowing for imaging of a whole range of different tissues.  Furthermore MRI is non-invasive and does not expose the subject to dangerous ionizing radiation, making it largely risk free, although the large magnetic fields involved exclude its use to patients with some implants. MRI is widely used to image all parts of the body for anatomical determinations such as detecting tumors and brain imaging. In the following we outline the main principles of the technique. Further details on the wide range of imaging modes are provided in the dedicated literature~\cite{QMMRI, MRIprimer}.

\subsection{Nuclear Magnetic Resonance}
~\index{nuclear magnetic resonance}~\index{nuclear magnetic resonance}
\label{NMR} A nucleus with non-zero spin, such as hydrogen, in an external magnetic field along the $z$ axis, $B_z$, will experience a Zeeman splitting of the otherwise degenerate energy levels corresponding to different $z$ components of the nuclear angular momentum~\cite{Blundell}. The result is a spontaneous polarization of the system, however, since the energy splitting is very low compared with typical thermal energies, the resulting polarization is very small. In order to have an appreciable signal very strong magnetic fields are required, which tend to be of the order 0.3--2\,T in the used instrumentation although they can be much higher. Manipulation of the resulting magnetization is now possible through the application of suitable radio frequency (RF) field pulses which need to be set around the resonant frequency of the system for appreciable effects. This frequency is known as the Larmor frequency: $\omega_L = \gamma B_z$ where $\gamma$ is the gyromagnetic ratio.~\index{Larmor frequency} The large natural abundance of $^1$H in a human body, in combination with its comparatively large gyromagnetic ratio, make hydrogen an attractive choice for imaging biological systems, however, other nuclei such as phosphorous or sodium are also sometimes chosen. Since the Larmor frequency is also dependent on the chemical environment of the proton, a so-called chemical-shift, it is possible to gain contrast based on the unique fingerprint of particular molecules or on the general environment of the nucleus. In the initialized state the magnetization is aligned with the strong external field. The RF field pulse is then applied to rotate this magnetization away from the $+z$ axis to an extent which is determined by the intensity and duration of the pulse. Following the removal of the pulse, the magnetization will decay back to the initial state. The relaxation dynamics of the magnetization back to its equilibrium value, $M_0$, are described by the Bloch equations:

\begin{eqnarray}~\index{Bloch equation}
\label{eq:Bloch}
\frac{dM_x}{dt}&=\gamma (\vec{M} \times \vec{B})_x&-\frac{M_{x}}{T_2},\\
\frac{dM_y}{dt}&=\gamma (\vec{M} \times \vec{B})_y&-\frac{M_{y}}{T_2},\\
\frac{dM_z}{dt}&=\gamma (\vec{M} \times \vec{B})_z&-\frac{M_z-M_0}{T_1}.
\end{eqnarray} The first term in each equation describes the precession of the magnetization around the field. The second term in the first two equations describes the relaxation of the transverse magnetization over a characteristic timescale known as the $T_2$ relaxation time, which has contributions due to spin dephasing as a result of inhomogeneities in the local effective field and also spin-spin relaxation processes. The second term in the last equation describes the relaxation of the longitudinal magnetization and involves the dissipation of the excitation energy of the system back to the lattice over a time characterized by the spin-lattice relaxation time, $T_1$. By changing the measurement scheme the relative contribution of $T_1$ and $T_2$ to the relaxation can be varied and since these values are also influenced by local environments they provide a flexible range of contrast conditions for the resulting images in addition to the basic proton density contrast. $T_1$, for example, generally increases with the strength of the applied field whereas the $T_2$ values are relatively insensitive to this and hence field cycling experiments where the strength of the uniform field is changed allow one to investigate this dependence. Furthermore, by setting the duration of the excitation pulse, the initial state can be changed. For example, with a so-called $\pi/2$ pulse the magnetization is nutated into the $x-y$ plane whereas with a $\pi$ pulse the magnetization rotates to the $-z$ axis, in which case no precession of the magnetization occurs and the decay is determined by the $T_1$ time. If the magnetization is subsequently rotated back to the $x-y$ plane the initial amplitude provides contrast weighted by $T_1$. In all cases the precession and relaxation of the magnetization leads to a changing magnetic flux which can be detected as an induced voltage in a series of pick-up coils outside of the imaged object. The same coils can be used for the RF field generation and the signal detection. In some cases, if no suitable contrast can be obtained from the naturally occurring differences in tissue within the body, contrast enhancing agents may be administered, e.g. intravenously. These are typically paramagnetic gadolinium compounds or superparamagnetic iron oxide nanoparticles which modify the magnetic environment of the imaging region. 

\subsubsection{Imaging and Pulse Sequences}
In order to build up an image using NMR it is necessary to encode the spatial information about the region of signal generation in the global signal detected by the pick-up coils. This is achieved by using time-varying field gradients of the order of 10\,mT/m which are superimposed on the uniform global field in order to provide spatially varying Larmor frequencies. A variety of schemes exist to exploit this in imaging. For example, if a gradient field along the $z$ axis is applied during the initial excitation, only a particular slice will be in resonance with the RF pulse and hence this slice is selectively probed in the measurement. The thickness of the slice is determined by the gradient of the field and the bandwidth of the excitation pulse. For spatial localization within the slice, $x-y$ field gradients are subsequently applied and a Fourier approach is used to reconstruct the real-space image.  Due to the resulting different precession frequencies for the different regions of the slice, the nuclear moments, which are initially in-phase, begin to de-phase and hence the signal as a function of time progressively represents different spatial frequencies of the image. By appropriately varying the pulses, field gradients and the time delay between excitation and measurement, the whole of $k$-space can be probed and by mathematical transformation the real-space image can be extracted.

Depending on the measurement scheme, a very wide variety of field pulse sequences are applied to the sample which vary the weighting of the measurement to the different relaxation times and vary the order in which $k$-space is sampled. A typical sequence may start with the RF pulse during the application of the $z$-gradient field in order to select a $z$-slice. This is then followed by an incrementing y-gradient field with a subsequent $x$-gradient field applied for each step such that $k$-space is sampled in a Cartesian scheme. More advanced schemes, however, apply the incremental $x$ and $y$ gradients concurrently in order to map out radial or spiral $k$-space trajectories which can be more efficient, reducing the time required for data acquisition and enabling imaging of dynamic processes. 
~\index{spin echo}
One further protocol can be employed in order to counteract the dephasing caused by non-time varying spatial gradients in the global field and thereby separate this contribution to the $T_2$ relaxation time. One approach is known as a spin-echo technique. In the first stage a $\pi/2$ pulse is applied to nutate the magnetization into the $x-y$ plane. The system is then allowed to precess for a certain time, $\tau$, during which the spins will dephase due to the spatially inhomogeneous local fields. Next a $\pi$ pulse is applied, after which the phase differences between the spins will have been reversed. Further precession will gradually bring the spins back to their original in-phase state after a time $2\tau$. In this manner the influence of field inhomogeneities and the spread in fields due to chemical shifts are corrected for and the measurement is able to probe the spin-spin interactions from neighbouring nuclei.

\subsubsection{Functional Magnetic Resonance Imaging}
~\index{functional MRI}
Functional magnetic resonance imaging (fMRI) is one of the newer imaging modes which is employed to investigate activity in the brain. It relies on the change in the magnetic properties of blood cells depending on their oxygenation state. Oxygen in the blood is carried by the protein hemoglobin which is paramagnetic in its deoxygenated state. This paramagnetism modifies the local field felt by nearby water molecules, thereby impacting the effective $T_2$ values in the vicinity of blood vessels carrying deoxygenated blood as compared to those carrying oxygenated blood. This contrast mechanism is known as blood-oxygen-level dependent or BOLD. This can be employed for functional brain imaging since neural activity in a certain region is accompanied by a spike in the delivery of oxygenated blood to that region.

\subsection{Nuclear Quadrupole Resonance Imaging}
~\index{nuclear quadrupole resonance imaging}
Nuclear Quadrupole Resonance (NQR) is a related technique to NMR which can be used for imaging in an analogous manner to MRI
~\cite{NQRch}. Many of the underlying concepts of signal and image generation, as well as the required instrumentation, are identical. The main difference is that the technique does not employ the strong uniform background field to provide the energy splitting of nuclear states but rather the splitting is determined by the interaction of nuclear quadrupole electric moments and the electric field gradients around the nucleus. Such field gradients are determined by the precise chemical environment of the nucleus and as such the energy splitting and associated resonant frequencies can be used for chemical fingerprinting. NQR is restricted to nuclei with a spin quantum number $\ge1$ for which the quadrupole moment is non zero. As with NMR, the system is excited using RF field pulses and the relaxation of the system is detected via the free induction decay signal generated in pick-up coils. Imaging is achieved in a similar manner to MRI by applying spatially varying RF fields or magnetic fields~\cite{NQR1}. Field cycling NQR systems can be employed to improve the sensitivity of the measurements, for example in the case of low abundance of the investigated nuclei, in an approach combining NQR with NMR. In a typical measurement the field at the sample is alternated between a high field and low or zero field condition. This is most readily achieved by physically moving the sample into and out of a field region. In the high field environment, the large nuclear splitting occurs, polarizing the nuclei. The sample is then transferred to a low field region and the RF pulse applied at the appropriate NQR frequency to excite the state. Finally, the sample is transferred back to the strong field and an NMR measurement is carried out. In addition to the chemical environment, the resonance lines are strongly affected by physical parameters of the system and hence the technique has also been employed for imaging temperature, stress and pressure in a sample~\cite{NQR2}.
\subsection{Magnetoencephalography}
~\index{magnetoencephalography}
\label{sec:MEG}
Magnetoencephalography~\cite{MEG1, MEG2, MEG3} (MEG) is an emerging brain imaging technique which senses brain functionality by detecting the magnetic fields outside of the brain generated due to the ion currents associated with neuronal activity.  However, due to the extremely weak signals which are of the order 10-100\,fT, particularly sensitive magnetometers are required to detect the generated fields and even then the detected signals necessarily correspond to the simultaneous firing of thousands of neurons from a small volume. The principal magnetometer of choice is a dc SQUID which is an incredibly sensitive device for magnetic field detection. ~\index{SQUID}A typical MEG setup consists of an array of hundreds of SQUIDs arranged in a grid so as to detect the field over the whole scalp. The SQUID loops can be wound in different configurations in order to directly detect the field strength or by using either multiple stacked coils or variously twisted loops the out-of-plane or in-plane field gradients can also be detected. Due to the need to operate the superconducting elements at either liquid helium or liquid nitrogen temperatures, the whole array sits at the bottom of a cryostat, the base of which is concave so as to be better moulded to the head of the patient who sits underneath. The very small signals being measured require that the whole apparatus needs to be housed in a magnetically shielded room, or else compensation procedures need to be applied in order to correct for distant magnetic field sources which would otherwise swamp the signal of interest. Furthermore care needs to be taken during measurement to distinguish the brain signal from artifacts from fields originating from elsewhere in the body such as from the heart or eye regions.

MEG has a number of particular advantages in brain imaging studies. Since the technique is completely non-invasive it is a particularly safe tool. The temporal resolution of the technique is also competitive, being able to detect changes on a sub-ms timescale. It can be seen as a complementary imaging technique to the related electroencephalography (EEG) which detects the concurrently generated electric fields, since both techniques are sensitive to differently oriented current dipoles within the cranium. MEG also tends to have a better spatial resolution than EEG, down to a few mm, and furthermore it is less sensitive to the conductivity variations from the detailed structure of the head which distort the EEG signal. Particularly active areas of application include studies of epilepsy and autism.

\subsubsection{The Inverse Problem}
Given a known current distribution and a knowledge of the details of the surrounding medium such as the geometry, electric permittivity and magnetic permeability, it is a relatively straightforward exercise to calculate the resulting electromagnetic field distributions from Maxwell's equations. In an MEG measurement the task is to calculate the current distributions that were responsible for generating the measured field. Unfortunately it has been shown that there is no unique solution to this so-called inverse problem and hence progress requires the development of simplified models of the system which are suitably constrained to yield a physically relevant solution. The simplest descriptions of the current distributions approximate the current sources as current dipoles in an equivalent dipole model. An equivalent current dipole represents a spatial average of the source currents within a small volume of the brain, characterized by its position, orientation and strength. Depending on the complexity of the model, different numbers of equivalent current dipoles can be assumed in order to fit the data. For a quasi-continuous model the brain is divided into a large number of discrete volume cells known as voxels and each voxel is allocated 3 orthogonal equivalent current dipoles. The task is then to determine the strength of each equivalent current dipole in order to recreate the observed field distribution. Whichever model is used, the parameters are iteratively adjusted and the resulting field is calculated in order to minimize the deviation between the model and the experimental data. However, depending on the model chosen the problem can be severely underdetermined and in any case due to the lack of a unique solution it is necessary to sensibly constrain the problem. Models of the head usually approximate it as a uniform spherical conductor. Improved modelling and constraints to the fitting algorithms can be provided by complementary MRI imaging of the brain to provide accurate anatomical details.

\section{Summary}
\label{sec:summary}

\begin{table}[!ht]
\centering
\caption[]{Comparison of magnetic imaging techniques, presenting some of the key specifications and attributes.  The quoted values are in general typical achievable values.$^*$Proof of concept recently demonstrated~\cite{timeSEMPA, timeSEMPA2, timeSPSTM}}

\begin{tabular}{|C{1.75cm}||C{1.5cm}|C{1.55cm}|C{1.55cm}|C{1.5cm}| C{3.3cm} |}
\hline\noalign{}
\textbf{Technique} & Probed Quantity & Spatial  Resolution& Temporal Resolution & Info. Depth& Comments \\

\hline\noalign{}

Lorentz Microscopy & stray field \newline + \newline sample induction& 10\,nm&1\,ns&sample average&Thin samples, \newline Quantitative info. with differential phase contrast microscopy.\\
\hline\noalign{}
Electron Holography&  stray field \newline + \newline sample induction&5\,nm &10\,ms & sample average&Quantitative info. through mathematical image reconstruction. \\
\hline\noalign{}
SEMPA& magnetization&20\,nm &700\,ps$^*$& 1\,nm&Quantitative info.,\newline Long acquisitions, \newline UHV required.  \\
\hline\noalign{}
SP-STM& magnetization &atomic&120\,ps$^*$& surface & UHV required, \newline Usually low temperature, \newline Long acquisitions. \\
\hline\noalign{}
MFM& stray field &10-100\,nm&low& 1000\,nm& Potentially invasive, \newline Long acquisitions, \newline Few sample requirements. 
\\
\hline\noalign{}
TXM&magnetization&25\,nm &50\,ps& sample average&Synchrotron technique, \newline Quick overview images.
\\
\hline\noalign{}
STXM&magnetization&25\,nm &50\,ps& sample average&Synchrotron technique,  \newline High repetition rates. 
\\
\hline\noalign{}
PEEM&magnetization&25\,nm &50\,ps& 5\,nm& Synchrotron technique, \newline Discharges possible due to high potential.
\\
\hline\noalign{}
CDI&magnetization&40\,nm &fs-ps& sample average&Zero drift, \newline Synchrotron technique \newline Complex sample fabrication $\&$ image reconstruction.
\\
\hline\noalign{}
MRI&proton density\newline $\&$ environment&1-2\,mm&100\,ms-several \newline \,sec.&3D imaging&Low risk, \newline Very versatile.
\\
\hline\noalign{}
MEG&stray field&5\,mm&$<$1\,ms&3D imaging via modelling&No unique solution, \newline Risk free.
\\
\hline
\end{tabular}
\label{tab:comp}       
\end{table}

As has been presented, a wide variety of techniques are available to image the magnetic state of a system and for a given application it is necessary for the user to judge the most appropriate option depending on the type of specimen, the information that one wants to acquire and the required spatial and temporal resolution. To conclude, table~\ref{tab:comp} provides a summary of some of the key attributes of a selection of the most widely employed techniques to enable ease of comparison. We note that the quoted values are not necessarily the ultimate limits of the techniques, but rather in most cases represent typical values under standard operating conditions.

\section*{Acknowledgements}
\label{sec:ackowledgements}
\addcontentsline{toc}{section}{Acknowledgements}
A large number of students, postdocs, colleagues and collaborators have also been involved in the authors' research efforts on magnetic imaging over the years, only a few examples of which we have been able to present here. Without these individuals, this work would not have been possible and we gratefully acknowledge all their contributions and insights.

\newpage
%
%

%
%

\end{document}